\documentclass[usenatbib,useAMS]{mnras}

\usepackage{amssymb}	

\usepackage{newtxtext,newtxmath}

\usepackage[T1]{fontenc}
\usepackage[normalem]{ulem}

\DeclareRobustCommand{\VAN}[3]{#2}
\let\VANthebibliography\thebibliography
\def\thebibliography{\DeclareRobustCommand{\VAN}[3]{##3}\VANthebibliography}

\usepackage{graphicx}	
\usepackage{amsmath}	

\usepackage{fontawesome5}
\usepackage{color}
\usepackage{xcolor}

\newcommand{\Rvir}{R_{\rm vir}}
\newcommand{\Mvir}{M_{\rm vir}}
\newcommand{\cvir}{c_{\rm vir}}

\newcommand{\eg}{{\sl e.g.}, }     
\newcommand{\dex}{{\,\,\rm dex}}

\newcommand{\hinv}{\ensuremath{\, h^{-1}}}%
\newcommand{\msol}{\ensuremath{\, {\rm M}_\odot}}    
\newcommand{\msun}{\ensuremath{\, {\rm M}_\odot}} 
\newcommand{\kpc}{\ensuremath{\, {\rm kpc}}}         
\newcommand{\mpc}{\ensuremath{\, {\rm Mpc}}}

\definecolor{orcidlogocol}{HTML}{A6CE39}
\definecolor{purple}{RGB}{128, 0, 128}

\newcommand{\OrcidID}[1]{ \href[urlcolor = red]{https://orcid.org/#1}{\textcolor{lightgray}{\faOrcid}}}
\newcommand{\OrcidIDName}[2]{\href{https://orcid.org/#1}{#2}}

\defcitealias{Anbajagane2022BaryImprint}{A22}

\usepackage[T1]{fontenc}
\usepackage{ae,aecompl}

\usepackage{newtxtext,newtxmath}

\title[Baryonic imprints on halo concentration]{Baryonic Imprints on DM Halos: The concentration--mass relation in the CAMELS simulations}

\author[Shao, Anbajagane \& Chang]{
\OrcidIDName{0000-0002-4561-7026}{Mufan (Jon) Shao},$^{1}$\thanks{Author emails: mufan@uchicago.edu, dhayaa@uchicago.edu}
\OrcidIDName{0000-0003-3312-909X}{Dhayaa Anbajagane},$^{2,\,3\star}$
\OrcidIDName{0000-0002-7887-0896}{Chihway Chang}$^{2,\,3}$\\
\\
$^{1}$ Department of Physics, University of Chicago, Chicago, IL 60637, USA\\
$^{2}$ Department of Astronomy and Astrophysics, University of Chicago, Chicago, IL 60637, USA\\
$^{3}$ Kavli Institute for Cosmological Physics, University of Chicago, Chicago, IL 60637, USA}

\date{Accepted XXX. Received YYY; in original form ZZZ}

\pubyear{2022}

\begin{document}
\label{firstpage}
\pagerange{\pageref{firstpage}--\pageref{lastpage}}
\maketitle

\begin{abstract}
The physics of baryons in halos, and their subsequent influence on the total matter phase space, has a rich phenomenology and must be well understood in order to pursue a vast set of questions in both cosmology and astrophysics. We use the \textsc{Camels} simulation suite to quantify the impact of four different galaxy formation parameters/processes (as well as two cosmological parameters) on the concentration--mass relation, $\cvir-\Mvir$. We construct a simulation-informed nonlinear model for concentration as a function of halo mass, redshift, and 6 cosmological/astrophysical parameters. This is done for two galaxy formation models, \textsc{IllustrisTNG} and \textsc{Simba}, using 1000 simulations of each. We extract the imprints of galaxy formation across a wide range in mass $\Mvir \in [10^{11}, 10^{14.5}] \msun/h$ and in redshift $z \in [0,6]$ finding many strong mass- and redshift-dependent features. Comparisons between the \textsc{IllustrisTNG} and \textsc{Simba} results show the astrophysical model choices cause significant differences in the mass and redshift dependence of these baryon imprints. Finally, we use existing observational measurements of $\cvir - \Mvir$ to provide rough limits on the four astrophysical parameters. Our nonlinear model is made publicly available and can be used to include \textsc{Camels}-based baryon imprints in any halo model-based analysis.
\end{abstract}




\section{Introduction}

The distribution of matter in our Universe contains information on both the initial conditions of the Universe, i.e. the primordial density field, and the subsequent cosmological and astrophysical evolution of the field to late times \citep[\eg][]{Allen2011CosmoClusterReview, Weinberg:2013review, Kravtsov2012ClusterFormation}. Many efforts have been made in the past decades to measure this structure and use it to understand processes of both cosmological and/or astrophysical nature \citep[\eg][]{Asgari2021KidsWL, Pandey2021DESxACT, Gatti2021DESxACT, Secco2022Y3Shear, Amon2022Y3shear, DES2022Y3, Troster2022KidstSZ, Schneider2022KidsAstro} 

Efforts to understand the astrophysical impact in particular have grown significantly in recent times, especially as it becomes clearer that our lack of knowledge on these astrophysical processes will impact our ability to infer precise and unbiased cosmological constraints from future measurements \citep{Sembolini2011BaryonsWL, Chisari2018BaryonsPk}. The impact of baryons on structure formation is strongest on ``nonlinear'' length scales corresponding to the sizes of massive halos, and thus is intimately connected to galaxy formation processes \citep[\eg][]{Rudd2008BaryonsMPk, Schneider2019BaryonsPk}. The influence of such non-gravitational processes on the baryon matter distribution is indirectly felt by the dark matter (DM) distribution via the gravitational coupling between the two components.

These astrophysical impacts on the matter distribution --- either directly in the case of gas and stellar distributions, and indirectly in the case of the DM distribution --- will be reflected in the density profiles of the halos. These profiles generally follow an NFW form \citep{Navarro1997NFWProfile}, where for a halo of a given mass the only free parameter is the halo concentration, $\cvir$. The changes in the matter distribution of a halo due to astrophysics will result in changes in its $\cvir$ value.

On nonlinear scales, many aspects of the observed structure can be understood and modelled using these profiles of halos \citep{Cooray2002HaloModel}; in recent times such halo-based approaches have been used to model the nonlinear galaxy and shear fields \citep[\eg][]{Zacharegkas2021GGLensingDES}, the outskirts of halos \citep[\eg][]{Shin2021MassGalaxyProfilesDES, Anbajagane2022Shocks}, and the cross-correlation between matter and thermodynamic fields \citep[\eg][]{Vikram2017GalaxyGroupstSZ, Pandey2019GalaxytSZ, Pandey2021DESxACT, Sanchez2022Sheary}. Thus, the question of understanding astrophysical impacts on nonlinear structure formation can be reformulated into understanding such impacts on the halo concentration, or more generally its impacts on the halo density profile \citep[\eg][]{Schneider2019BaryonsPk}.

The impact of astrophysics on halo concentration has a long history of study, first performed in smaller samples \citep[\eg][]{Gnedin2004AdiabaticContraction, Bryan2013BaryonImpactOnShapes}, and more recently in samples numbering millions of halos and spanning multiple decades in halo mass \citep{Ragagnin2019HaloConcentration, Beltz-Mohrmann2021BaryonImpactTNG, Anbajagane2022BaryImprint}. A caveat however is that different hydrodynamic simulations employ varying prescriptions for galaxy formation, including the actual equations that are solved in the simulation. While this variety in prescriptions has been useful in setting bounds on the impact of astrophysics on halo properties, it does not enable a systematic study of astrophysics in halos. There is indication that the form of the mass-dependent impact of baryons on the halo matter distribution is similar across state-of-the-art galaxy formation models but the quantitative behavior is still different \citep{Chua2021TNGShapes, Anbajagane2022GalVelBias}.

The \textsc{Camels} suite of hydrodynamic simulations \citep[][]{Villaescusa-Navarro2021CAMELS} was designed for quantitative studies of astrophysical impacts on structure formation. The suite contains $\mathcal{O}(10^3)$ simulations of a relatively small volume of halos while also varying four astrophysical simulation parameters --- two corresponding to supernovae (SN) feedback, and two corresponding to active galactic nuclei (AGN) feedback --- in addition to the two cosmological parameters most relevant for structure formation; $\Omega_{\rm m}$, the fraction of matter energy density at $z = 0$, and $\sigma_8$, the root-mean-squared fluctuation of the $z = 0$ density field smoothed on the $8 \mpc\hinv$ length scale. These simulations are also run with two different galaxy formation models, \textsc{IllustrisTNG} and \textsc{Simba}, which then provides a way to estimate model-dependent differences in the emergent astrophysical impacts.

A limitation of the \textsc{Camels} suite, however, is its small box size of $L = 25 \mpc/h$, which does not contain many halos of group and cluster mass-scales ($\Mvir \sim 10^{13} - 10^{14} \msol/h$). The astrophysical impacts on halos of these mass-scales are known to be distinct from the impacts in less massive counterparts, so accurately capturing these effects at high mass is of interest to this study. In addition, such massive halos have a notable impact on the observed structure in our Universe given they are large halos that form in the most overdense regions, even though their number counts are far lower than their less massive counterparts. 

In this work, we combine 1000 \textsc{Camels} simulations per galaxy formation model to create an ``uber'' halo sample. This sample is then used to build a non-parameteric model, constructed using a local linear regression technique, that can predict the concentration, $\cvir(\Mvir, z, \Omega_{\rm m}, \sigma_8, A_{\rm SN1}, A_{\rm AGN1}, A_{\rm SN2}, A_{\rm AGN2})$, which is a function of halo mass, redshift, two SN feedback parameters, two AGN feedback parameters, and two cosmological parameters. The choice of a local linear regression allows us to capture non-linear behaviors with halo mass while still using the familiar concept of power-law scaling relations. Such non-linearity with mass is an expected feature in galaxy formation given the many different mass scales of different processes, and this has been shown in various works across wide mass ranges \citep[\eg][]{Farahi2018BAHAMAS, Chua2019ShapeIllustrisBaryons, Anbajagane2020StellarStatistics, Chua2021TNGShapes, Anbajagane2022BaryImprint, Anbajagane2022GalVelBias}.

Our halo concentration model is then examined to understand the mass-dependent impacts of astrophysical feedback on the $\cvir-\Mvir$ relation; specifically to identify certain features/trends in the relation and motivate the physical processes that lead to such behavior. We do this over a wide range in redshift, spanning from $z \sim 3$, near the peak of star formation, all the way to the present epoch. While we generally will discuss results for both the \textsc{IllustrisTNG} and \textsc{Simba} models, our more detailed discussions will lean towards the \textsc{IllustrisTNG} results as we use the existing work of \citet*[henceforth referred to as \citetalias{Anbajagane2022BaryImprint}]{Anbajagane2022BaryImprint} on baryon imprints in the fiducial \textsc{IllustrisTNG} simulations to identify the different physical processes that may underlie the features found in our results here. In fact, our work is an extension on an aspect of \citetalias{Anbajagane2022BaryImprint}, where we now focus on just the concentration--mass relation but use \textsc{Camels} to extract the quantitative impact of galaxy formation on this relation.

To summarize, our goals in this work are to (i) construct and examine a model of the concentration as a function of astrophysical processes, with the dependence on redshift, halo mass, and cosmology folded in, (ii) to compare and contrast results from different formation models provided by the \textsc{Camels} suite (\textsc{IllustrisTNG} and \textsc{Simba}), and (iii) provide qualitative, observational constraints on the astrophysical \textsc{Camels} simulation parameters using available data on the $\cvir-\Mvir$ relation.

This work is organized as follows. We present our simulation data and the local linear regression method in \S \ref{sec:Data&Method}. We showcase our model, including the mass- and redshift-dependent impact of astrophysics, and the difference between \textsc{IllustrisTNG} and \textsc{Simba}, in \S \ref{sec:SimulationResults}. We then discusses uses of our model, including simulation ``translations'' and observational constraints, in \S \ref{sec:ModelApplications}. We conclude and discuss future extensions of this work in \S \ref{sec:Conclusion}.

\section{Data \& Methods}\label{sec:Data&Method}

\subsection{CAMELS Simulation Suite}\label{sec:Camels}

\begin{figure}
    \centering
    \includegraphics[width = \columnwidth]{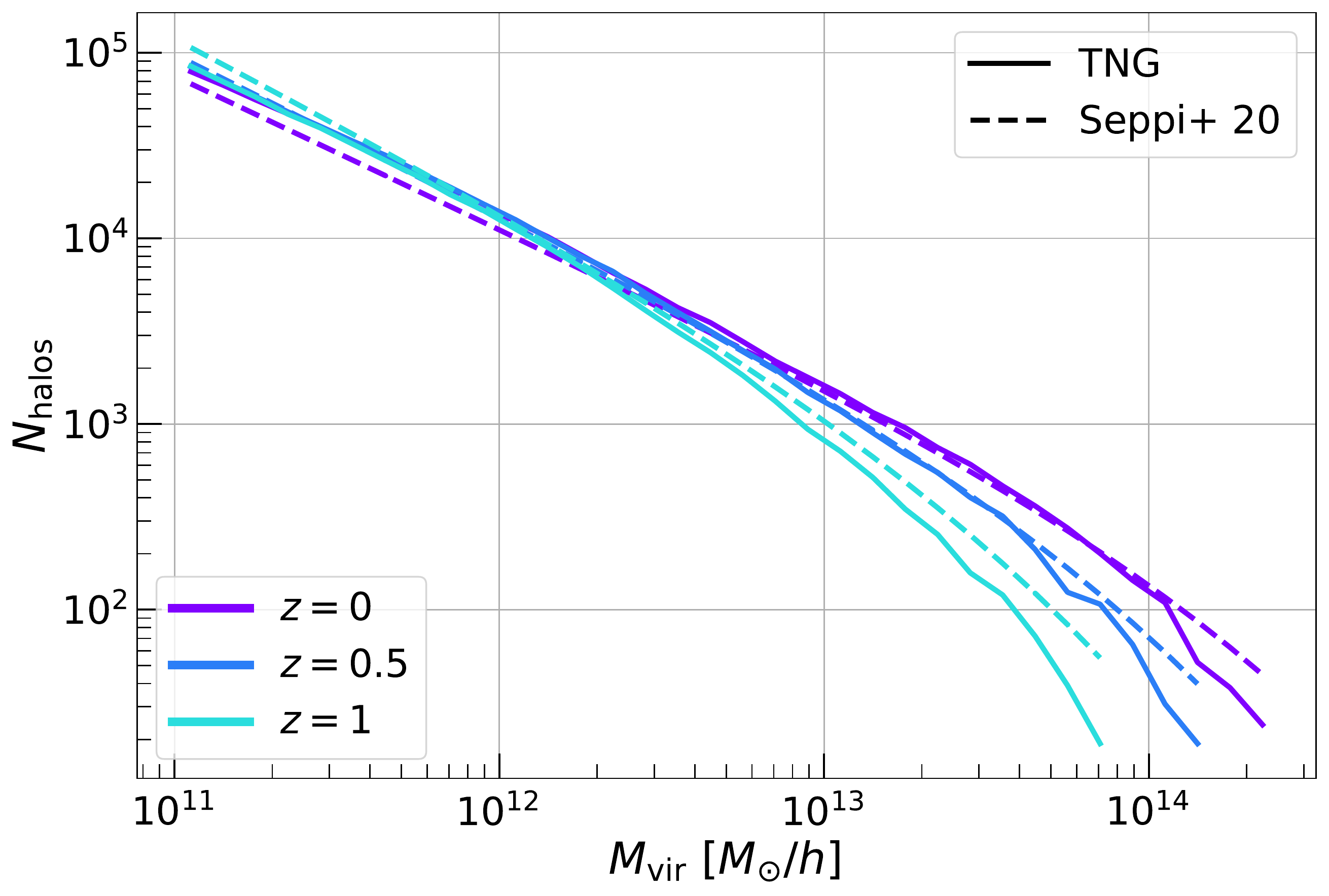}
    \caption{The halo counts as a function of halo mass for the uber sample of 1000 \textsc{Camels} simulations at three redshifts $z \in \{0, 0.5, 1\}$. We only show results for the \textsc{IllustrisTNG} formation model. The halo counts are computed in bins of $0.1 \dex$. The dashed lines show the theoretical expectation using the halo mass function model of \citet{Seppi2021HMF} and the theory accounts for the 1000 different cosmologies combined in the uber sample. The measured counts generally agree with the theory, validating that the uber sample is still a cosmologically consistent sample.}
    \label{fig:HaloCounts}
\end{figure}

The Cosmology and Astrophysics with MachinE Learning Simulations \citep[\textsc{Camels},][]{Villaescusa-Navarro2021CAMELS} is a simulation suite of $V = (25 \mpc/h)^3$ boxes with varying cosmological and astrophysical parameters, and multiple galaxy formation models. The specific cosmological parameters are $\Omega_{\rm m}$ and $\sigma_8$, which are the two parameters that have the strongest influence on structure formation, and the astrophysical parameters are $A_{\rm SN1}, A_{\rm SN2}, B_{\rm SN1}, B_{\rm SN2}$ controlling supernovae feedback, and $A_{\rm AGN1}, A_{\rm AGN2}, B_{\rm AGN1}, B_{\rm AGN2}$ which control active galactic nuclei (AGN) feedback. The specific meaning of each parameter differs based on the chosen galaxy formation model --- \textsc{IllustrisTNG} or \textsc{Simba} --- and these differences are detailed further below in \S \ref{sec:TNGDescription} and \S \ref{sec:SIMBADescription} respectively. To highlight these differences explicitly, we will refer to these parameters with different notations, as $A_X$ for \textsc{IllustrisTNG} and $B_X$ for \textsc{Simba} respectively.

The \textsc{Camels} simulations were run with $256^3$ dark matter particles of mass $6.5\times 10^7 (\Omega_{\rm m} - \Omega_{\rm b})/0.251 \msol/h$, and then $256^3$ gas resolution elements with an initial mass of $1.27 \times 10^7 \msol/h$. The suite was designed specifically for training machine learning algorithms but the dataset is generally well-suited for non-parameteric studies of astrophysics and cosmology. There are different types of runs within the full suite, of which three are of interest to us. The \textbf{Cosmic Variance (CV)} runs are 27 simulations run with fiducial settings --- $\Omega_{\rm m} = 0.3$, $\sigma_8 = 0.8$, and $A_X = B_X = 1$, with $X \in \{\rm SN1, SN2, AGN1, AGN2\}$ --- and with varying initial conditions. The \textbf{1P} runs are 61 datasets (10 simulations per each of the 6 parameters and 1 fiducial run) that systematically vary a single parameter. All runs share the same initial conditions seed. Finally, the \textbf{Latin Hypercube (LH)} simulations are the main driver of this work. These are 1000 simulations that span a 6D parameter space of the input parameters of the simulation, with the points in parameter space chosen via an optimal scheme called a Latin Hypercube, and this sampling technique has become a frequent tool used in building emulators for cosmology \citep{Heitmann2009Emulator}. Note that in addition to the \textsc{IllustrisTNG} and \textsc{Simba} versions, \textsc{Camels} also has a dark matter only (DMO) version of every run (CV, 1P and LH). In this case, only the cosmological parameters matter and the astrophysical parameter values have no impact.

We make use of the \textsc{Rockstar} halo catalogs \citep{Behroozi2013Rockstar} from all these simulations. The halo concentration in these catalogs is computed by fitting the total matter density profiles of halos with an NFW form \citep{Navarro1997NFWProfile},
\begin{equation} \label{eqn:NFW_profile}
    \rho(r) = \frac{\rho_s}{\frac{r}{r_s}\big(1 + \frac{r}{r_s}\big)^2}\,,
\end{equation}
where $\rho_s$ and $r_s$ are the characteristic density and the scale radius, respectively. The concentration is then derived as,
\begin{equation}\label{eqn:c200c_def}
    \cvir = \Rvir/r_s,
\end{equation}
where the virial radius, $\Rvir$, is defined as the radius of the sphere containing an enclosed density that is $\rho = \Delta_{\rm vir}(z) \rho_c(z)$, where $\rho_c$ is the critical density at a given epoch, and $\Delta_{\rm vir}(z)$ is the density contrast described in \citet{Bryan1998vir}. The virial mass, $\Mvir$, is defined as the mass enclosed within this radius, $\Mvir = 4/3\pi\Rvir^3 \times \Delta_{\rm vir} \rho_c(z)$ and is the mass definition used throughout this work.

Note that the concentration and virial mass here are computed using \textit{all} available particles rather than just the DM alone. Thus, this work studies the NFW profiles of the \textit{total matter} component and not just the DM only. Though, given the DM constitutes 85\% of the matter content, the total matter distribution --- on length scales of interest to us, namely excluding the halo core where the stellar mass distribution is most prominent --- would to good approximation be given by just the DM distribution.\footnote{Note that \textsc{Rockstar} obtains $\cvir$ by fitting density profiles down to the force softening scale of the simulation, which is $\epsilon \approx2\kpc$ (comoving) for the simulations discussed here. In practice, the smallest informative length scale of the profile is 2-3 times the force softening as densities on scales close to $\epsilon$ are downweighted in the \textsc{Rockstar} fitting procedure.}
The total matter distribution is observationally more relevant, and is conveniently the definition used in these \textsc{Rockstar} catalog.

The total matter distribution has been shown to follow an NFW profile for $r > 0.1\Rvir$ but deviates from it below that radius scale due to the presence of the stellar component \citep[see their Figure 5]{Duffy2010BaryonDmProfileDensity}. The \textsc{Rockstar} fitting algorithm uses the measured profile down to a minimum scale set by the force softening scale of the simulation, $2 \rm kpc$. This is within $0.1\Rvir$ for all halos in our sample. So the \textsc{Rockstar} concentration estimates will be affected by the fact that the true density profile does not follow an NFW form on small-scales. Thus, we treat these estimates as NFW-based approximations of the true density profile.

We also perform some quality cuts on the raw catalogs. In particular, the \textsc{Rockstar} catalog shows a notable outlier halo population --- 2\% (5\%) at $z = 0$ ($z = 1$) --- that have $\cvir \lesssim 1$. We treat these objects as outliers and remove them from our analysis. In short, this is done by forming a histogram of $\cvir$, identifying where the low-$\cvir$ peak begins to rise and removing all halos below that limit. We detail this procedure more in Appendix \ref{sec:Outlier}.

The uber halo samples --- obtained by combining halos from all 1000 LH simulations --- span a mass range of $10^{11} \msun/h < \Mvir < 10^{14.5} \msun/h$; the minimum mass is set by requiring at least $\approx 1000$ particles in a halo, and the maximum mass is set by the number of available halos and varies by redshift. In all our analyses, we set the maximum $\Mvir$ for each redshift as the 10th most massive halo mass in the uber sample at that redshift. The distribution of masses for some redshifts is displayed in Figure \ref{fig:HaloCounts}, and shows that the uber sample follows a cosmological distribution of halos, as evidenced by the $\approx 20\%$ agreement between the measured counts and the theoretical predictions. The measurements come from aggregating counts from all 1000 LH simulations. The theory line is constructed by obtaining predictions for each of the 1000 individual simulations --- taken from the model of \citet{Seppi2021HMF} as implemented in \textsc{Colossus} \citep[][]{Diemer2018COLOSSUS} --- and aggregating them the same way we aggregate the counts in the LH simulations. In this way the theoretical prediction accounts for the cosmology variation across the 1000 LH simulations.

We now briefly detail the specific meaning of each astrophysical parameter in the two galaxy formation models, \textsc{IllustrisTNG} and \textsc{Simba}. More detail on the simulation models can be found in \citet[see their Section 3]{Villaescusa-Navarro2021CAMELS}.

\subsubsection{IllustrisTNG model}\label{sec:TNGDescription}

\textsc{IllustrisTNG}, which we will interchangeably refer to as TNG, is a state-of-the-art high-resolution suite of simulations \citep{Nelson2018FirstBimodality, Pillepich2018FirstGalaxies, Springel2018FirstClustering, Marinacci2018FirstFields, Naiman2018FirstEuropium, Nelson2019TNG50, Pillepich2019TNG50}. The same model and parameters from the fiducial simulations runs --- which are described in detail in \citet{ Weinberger2017Methods, Pillepich2018Methods} --- are adopted in the \textsc{Camels} simulations for this galaxy formation model. The resolution level of the \textsc{Camels} simulation roughly corresponds to the TNG100-2 run.

The total energy injection rate (power) per unit star-formation, $e_w$, and the SN wind speed, $v_w$, are modulated by two normalization factors, $A_{\rm SN1}$ and $A_{\rm SN2}$, respectively. Thus $A_{\rm SN1}$ controls the energy output rate, and $A_{\rm SN2}$ controls the wind speed. The mass loading factor, $\eta$, which sets how much mass is transferred away through the SN winds/feedback, is set by the combination $v_w^{-2}e_w$ and so is proportional to the combination $\eta \propto A_{\rm SN1}A_{\rm SN2}^{-2}$.

The supermassive black hole (SMBH) model has different modes of feedback --- a quasar/thermal mode, where nearby gas is thermally heated, and a radio/kinetic mode, where the gas is given a momentum kick. The kinetic mode is turned on during the lower accretion rate-phase of the SMBH, and is the dominant mechanism for halos at the Milky Way-scale and above \citep[\citetalias{Anbajagane2022BaryImprint}]{Weinberger2018SMBHsIllustrisTNG}. Additionally, the kinetic feedback is not a continuous output into the halo; instead the SMBH ``accumulates'' energy and expels it once an energy scale is reached, which results in a ``bursty'' feedback process. The parameter $A_{\rm AGN1}$ controls the rate at which energy is accumulated by the SMBH and $A_{\rm AGN2}$ controls the minimum energy needed to initiate such a burst. Both control the rate of the bursts, while the latter controls the speed of the kinetic feedback jets.

\subsubsection{SIMBA model}\label{sec:SIMBADescription}

\textsc{Simba} is another galaxy formation model \citep{Dave2019SIMBA} that contains a rich variety of astrophysical processes, such as SN, SMBH feedback etc. The implementation of the scaling parameters, $B_X$, in \textsc{Simba} is roughly similar to $A_{X}$ in \textsc{IllustrisTNG} but has some differences in the details of the implementation.

While $B_{\rm SN2}$ continues to correspond to the SN wind velocity like $A_{\rm SN2}$, $B_{\rm SN1}$ directly scales the mass-loading factor now (as opposed to in TNG where both SN parameters controlled the mass-loading). There are no parameters that directly scale the energy output of SN events. We stress that the specific equations governing the winds and the feedback still differ significantly between \textsc{IllustrisTNG} and \textsc{Simba}. Of particular note is that the \textsc{Simba} model is informed more strongly by results from existing ultra high-resolution galaxy-scale zoom-in simulations \citep{Muratov2015GasOutflowsFIRE}.

As for the SMBH model, the $B_{\rm AGN1}$ parameter in \textsc{Simba} sets the scale of the momenta kicks provided to the gas through the kinetic feedback mechanism, which is different from the energy accumulation rate that $A_{\rm AGN1}$ controls in TNG. Then, $B_{\rm AGN2}$ sets the velocity scale of the jets that result from the kinetic feedback, and this is similar to the role of $A_{\rm AGN2}$ in TNG. Now, once $B_{\rm AGN2}$ sets the jet velocity scale, the value of $B_{\rm AGN1}$ determines the mass in the AGN outflows that is required to obtain the momentum scale set by $B_{\rm AGN1}$.

\subsection{Observational $\cvir - \Mvir$ relation data}\label{sec:ObsData}

We use observational data for the $\cvir - \Mvir$ relation to place constraints on the astrophysical parameters in \textsc{Camels} ; namely to ask whether the constraints lie above/below the fiducial values. We take data mainly from two existing works --- \citet{Johnston2007Concentration, Mandelbaum2008ConcentrationMeasurement} --- both of which measured halo concentration from the Sloan Digital Sky Survey \citep[SDSS,][]{York2000SDSS}. The values for $\cvir$ were inferred from average lensing profiles of ensembles of halos. \citet{Mandelbaum2008ConcentrationMeasurement} impose scale cuts to extract parts of the profile that can be more easily compared to N-body simulations, while \citet{Johnston2007Concentration} used much smaller scales but accounted for the baryonic components when extracting the halo concentration. The estimates were made at various redshifts across works but can be brought to a common redshift of $z = 0$ using a simple scaling as was done in \citet{Child2018ConcentratioMassRelation}. We also convert the chosen overdensity definitions of the measurements, using the method of \citet{Hu2003SampleVar}, to match the virial mass definition of \citet{Bryan1998vir} which is the default output definition in the \textsc{Rockstar} halo catalog. This conversion is done by using the mean mass and concentration of a given datapoint to compute the offets to apply in mass and concentration and thus convert the definition. Notably, we do not incorporate any additional uncertainties on concentration and mass that arise from performing this procedure. This is adequate for the precision level of our analysis.

We also use three additional points --- one from \citet{Brimioulle2013CFHTCvir} at galaxy-scales using the Canada-France-Hawaii telescope (CFHT) data, and two points from \citet{Mandelbaum2006SDSSCvir} at group scales, using an older version of SDSS data than the ones used in \citet{Johnston2007Concentration, Mandelbaum2008ConcentrationMeasurement}. 

The two works of \citet{Johnston2007Concentration} and \citet{Mandelbaum2008ConcentrationMeasurement} both use SDSS data but have notable differences. They use different cluster catalogs that vary by $1.5 \dex$ in the mass range they cover. The latter also uses additional galaxy samples beyond those used by the former. There are furthermore many analysis choices that differ between the two in how they extract profiles, and how they account for different effects in the measurements. A more detailed description can be found in \citet[Section III. C]{Mandelbaum2008ConcentrationMeasurement}

When using this data to constrain model parameters, we must know the covariance of the data in order to compute the chi-squared, $\chi^2 = \delta x\mathcal{C}^{-1}\delta x$, where $\delta x$ is the residual between measurement and prediction. In our case, we assume our datapoints are uncorrelated, i.e. $\mathcal{C}$ is a diagonal matrix. This is certainly not true given some datapoints across different works have some shared data underlying them (notably, SDSS), but our limitation is the lack of any information on this point-to-point covariance. However, as we will stress later, our goal with using this observational data is not to make precision estimates of the parameter constraints but rather use the data to make qualitative statements, i.e. do the extracted parameter posteriors prefer values greater/lesser than the fiducial parameter values. For such a purpose, we find it suffices to approximate the covariance as a diagonal matrix, especially alongside other approximations above such as the mass definition conversion and the redshift re-scaling.

\subsection{Kernel-Localized Linear Regression}\label{sec:KLLR}

Traditionally, scaling relations are fit with a simple power law, i.e. a log-linear relation, or are extended to include a broken power-law model with two power-law indices instead of one. This is a perfectly sound technique, but does not capture all the non-linearity that can be present in the scaling relations. Using simple power-laws to represent scaling relations is well motivated in gravity-only simulations where there are very few unique scales in the problem that can break self-similar scaling. With the inclusion of galaxy formation, however, many scales now arise and lead to strong deviations from the power law behavior in the matter phase space \citepalias{Anbajagane2022BaryImprint}. Given the surplus of simulation data available now, one can explore models with more freedom that can accurately capture such non-linearities without assuming fixed functional forms.

A common model-independent approach has been to use machine learning (ML) techniques, such as random forest and neural networks \citep[\eg][]{Machado2020GasShapesSHAP, Stiskalek2022ML}. However, this has the issue of being difficult to interpret and extract physical explanations, though there are approaches one can take to make it easier \citep[\eg][\citetalias{Anbajagane2022BaryImprint}]{Ntampaka2019XrayClustersML, Machado2020GasShapesSHAP}. An alternative, is to go back to our well-known, well-understood linear models but extend them to capture some or most of the non-linearity.

Kernel-localized linear regression \citep*[\textsc{Kllr},][]{Farahi2022KLLR} is an extension that allows such flexibility. The input data is weighted according to a kernel, which for studying halo scaling relations is most often implemented as a gaussian kernel in $\log_{10}\Mvir$, and a linear regression is fit to the kernel-weighted data. As the kernel is moved over the log-mass direction, the method maps out the non-linear scaling relation using connected linear pieces. The technique has already been used extensively to show nonlinearity in different halo property--mass scaling relations \citep{Farahi2018BAHAMAS, Anbajagane2020StellarStatistics, Anbajagane2022BaryImprint, Anbajagane2022GalVelBias}. In this work, we use \textsc{Kllr} to study a multivariate scaling relation that is still modelled as being locally linear in halo mass; a version of \textsc{Kllr} that was first introduced and utilized in \citetalias{Anbajagane2022BaryImprint}. 

The locally linear, multivariate scaling relation of concentration is written in \textsc{Kllr} as,

\begin{align}\label{eqn:Kllr}
    \log_{10}\cvir = &\,\, \pi(\Mvir, z) + \alpha_M(\Mvir, z) \log_{10}\Mvir\nonumber\\
    & + \sum_X\alpha_X(\Mvir, z) \log_{10}X,
\end{align}x
where $\pi$ is the intercept, $\alpha_M$ is the slope with halo mass, $\alpha_X$ is the slope with other parameters, which are $X \in \{\Omega_{\rm m}, \sigma_8, Y_{\rm SN1}, Y_{\rm AGN1}, Y_{\rm SN2}, Y_{\rm AGN2}\}$, where $Y \in \{A, B\}$ depending on whether the model is for \textsc{IllustrisTNG} or \textsc{Simba}. The slopes and intercepts are all functions of halo mass and redshift. The concentration, $\cvir$, is thus modelled as a function of mass, redshift, and 6 cosmological/astrophysical parameters. The dependence on the cosmological/astrophysical parameters is notably also a function of mass. When we analyze the DMO simulations, we drop all astrophysical terms in equation \eqref{eqn:Kllr} and keep only the cosmology ones. When analyzing the CV simulations or any individual, single simulation, we drop all terms with $\alpha_X$ (since none of the parameters $X$ vary in these scenarios) and only keep the slope with mass.

We compute a \textsc{Kllr} model for each of the 34 available snapshots in $\textsc{Camels}$, using the uber sample at each redshift between $0 < z < 6$. We compute the mean $\cvir$ at every $0.1 \dex$ in mass, with the minimum mass set at $10^{11} \msun/h$ and a maximum mass set by the 10th most massive halo of the uber sample. The upper limit choice excludes regions where the model solutions are imprecise given the low number of halos (<10). The \textsc{Kllr} technique uses a Gaussian kernel in $\log_{10}\Mvir$, with a width of $0.3 \dex$.\footnote{In general, wider kernels smooth out more features, and narrower kernels lead to noisy estimates of the scaling relation. Our choice of $0.3\dex$ is guided by the semi-optimized choices of the previous works that used \textsc{Kllr}.} We obtain uncertainties on all the \textsc{Kllr} parameters of equation \eqref{eqn:Kllr} by obtaining the regression coefficients for 100 bootstrap realizations of the data. Thus, we have 100 versions of equation \eqref{eqn:Kllr} that can then be used to quantify the uncertainty in our \textsc{Kllr} model predictions.

Note that \textsc{Kllr} is by no means a unique method for capturing non-linearity in multivariate scaling relations via a method that is not an ML technique (random forest,  neural networks etc.). For example, one could utilize Gaussian processes to perform the regression instead; similar to \textsc{Kllr}, the Gaussian processes assumes no apriori information about the form of the scaling relation, and will capture non-linearities present in the data. Indeed, both \textsc{Kllr} and gaussian processes have already been used interchangeably to extract non-linear relations in halo quantities \citep[\eg][]{Farahi2021PoPE, Farahi2022ProfileCorr}.

\section{Results I: Baryon-dependent $\cvir-\Mvir$ relation}\label{sec:SimulationResults}

We present and examine the \textsc{Kllr}-based model built using the various \textsc{Camels} simulations. In particular, we introduce the different mean relations (\S \ref{sec:FidCvirMvirRelation}), detail the dependence of $\cvir - \Mvir$ on the astrophysical and cosmology parameters and the physical processes they originate from (\S \ref{sec:Astrophysics}), differences between TNG and SIMBA in the baryon imprints found in each (\S \ref{sec:TNGvsSIMBA}), the change with redshift of the astrophysical and cosmoogical dependence in $\cvir$ (\S \ref{sec:RedshiftDependence}), and finally the difference in behaviors in the hydrodynamic simulations with galaxy formation and their dark matter only counterparts (\S \ref{sec:TNGvsDMO}).

All results in this section come from our \textsc{Kllr} model results. Uncertainties are obtained from 100 bootstrap resamplings and represent the 68\% confidence interval.

\subsection{Fiducial $\cvir-\Mvir$ relation in \textsc{IllustrisTNG} and \textsc{Simba}}\label{sec:FidCvirMvirRelation}

\begin{figure}
    \centering
    \includegraphics[width = \columnwidth]{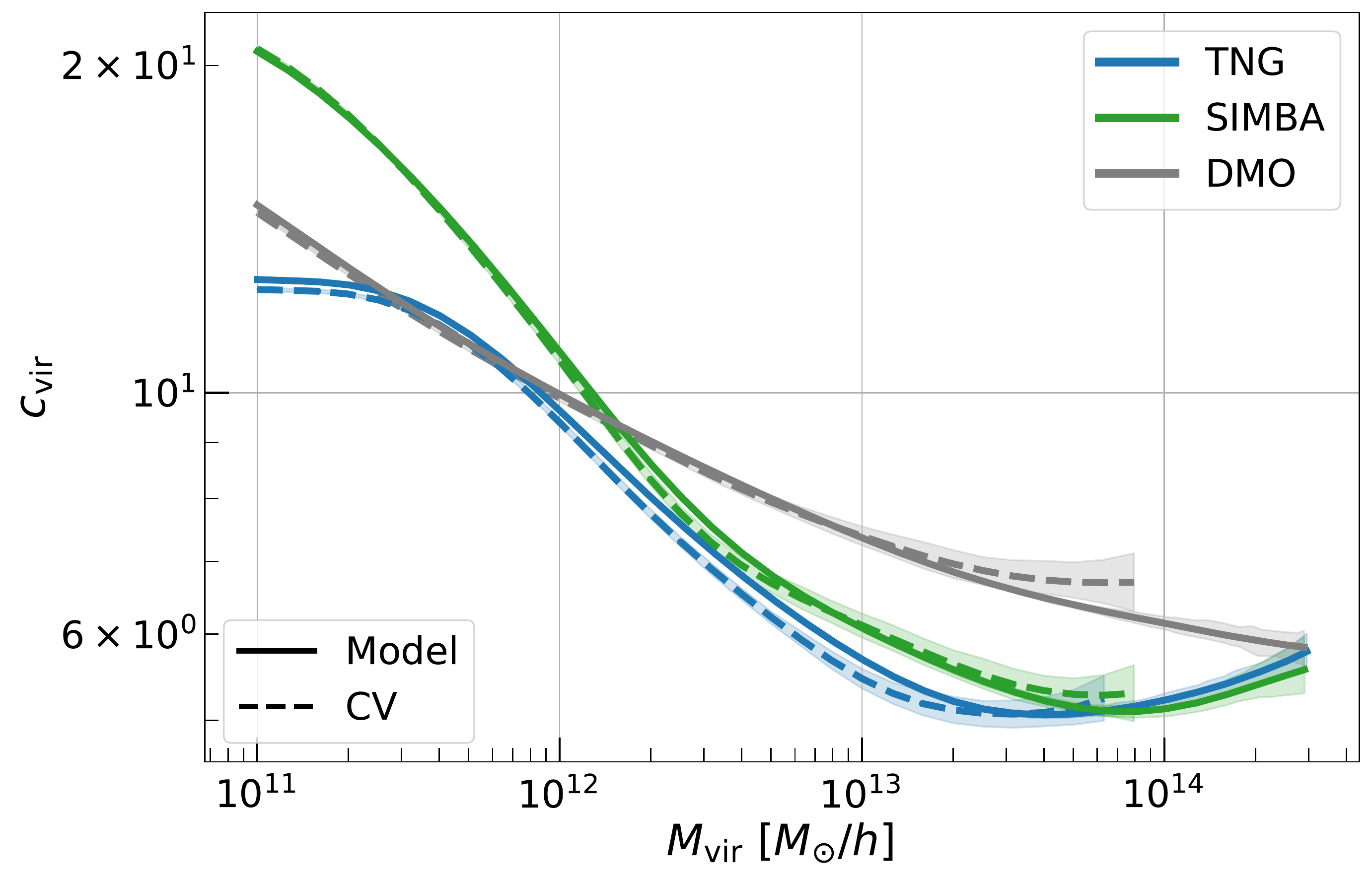}
    \caption{The $z = 0$ \textsc{Kllr} scaling relation of $\rm \cvir - \Mvir$ for TNG, SIMBA, and DMO runs. The bands show the 68\% confidence interval obtained from 100 bootstrap realizations for the data and for the model. The solid lines are predictions for the fiducial simulation parameters using our model that was constructued using the 1000 LH sims. The dashed lines are the \textsc{Kllr} scaling relations estimated on a combined halo sample from the 27 CV simulations. TNG and SIMBA agree well at higher halo masses, with both deviating significantly from DMO, and have bigger differences at low masses. In all cases, the prediction from the model agrees with the CV results.}
    \label{fig:TNG_SIMBA_DMO_CV}
\end{figure}

 Since the efficiency of dissipative processes peaks near MW masses, the concentration in the FP runs is boosted at that scale. However, competing feedback processes, from SN at lower masses and SMBHs at higher masses, disperse baryons out of the core and into the outer halo, thus decreasing the DM concentration at these mass scales. Above $\sim 10^{14}\msol$, the shift in mean concentration is very small. 

As mentioned before, the impact of baryons on halo concentration has a long history of study in simulations. The first discovery was that the presence of gas --- a component that could radiate away energy via cooling mechanisms --- led to adiabatic contraction, and this change in the gas distribution, through gravitational back-reaction on the DM component, caused halos to be more concentrated \citep{Blumenthal1986AdiabaticContraction, Gnedin2004AdiabaticContraction} and others have found the effect amplitude can depend on the history of how baryons assmebled in the halo \citep{Abadi2010ShapesBaryons, Tissera2010BaryonImprints, Pedrosa2010BaryonImprints, Artale2019BaryonImprints, ForouharMoreno2021BaryonsConcentrationEagle}. The subsequent implementation of stellar feedback, and then AGN feedback, in the simulations led to the discovery that the halo is made more diffuse through these explosive processes and this has been shown before for cluster-scale haloes \citep{Duffy2010BaryonDmProfileDensity, Cui2016NiftyBaryonsHaloProperties}, MW-scale haloes \citep{Pedrosa2010BaryonImprints, Artale2019BaryonImprints, ForouharMoreno2021BaryonsConcentrationEagle} and for dwarf galaxy-scale haloes \citep{Wetzel2016DwarfGals, Wheeler2019FIREdwarfs}. More recently, \citetalias{Anbajagane2022BaryImprint} showed the full mass-dependence of these effects in TNG across 6 decades in halo mass, from dwarf galaxies to galaxy clusters, and that these different effects interact to create a non-trivial ``wiggle'' feature in the concentration--mass relation. \citet{Anbajagane2022GalVelBias} showed --- using a different tracer of the halo potential --- that such mass-dependence from baryon imprints can be generally found in many of the latest hydrodynamic simulations, such as \textsc{IllustrisTNG}, \textsc{Magneticum}, \textsc{Bahamas} and \textsc{The300} \citep[][]{Nelson2018FirstBimodality, Pillepich2018FirstGalaxies, Springel2018FirstClustering, Marinacci2018FirstFields, Naiman2018FirstEuropium, Hirschmann2014MGTM, McCarthy2017BAHAMAS, Cui2018The300}. However, all of the existing works summarizing large samples of halos have made comparisons between three to four specific galaxy formation models. In this work, we have extended this to utilize $\mathcal{O}(10^3)$ variations of a given galaxy formation model, and we explicitly compare the impact of galaxy formation in a mass-dependent way.

We start in Figure \ref{fig:TNG_SIMBA_DMO_CV}, showing the comparison of several fiducial $\rm \cvir - \Mvir$ relations at $z = 0$; we show TNG (blue), SIMBA (green), and DMO (gray) versions. The solid lines are the predictions from our \textsc{Kllr}-based models --- which were developed using the 1000 LH simulations --- given fiducial values for the six input parameters of the simulations. The dashed lines are the \textsc{Kllr} scaling relation of a combined halo sample from all 27 CV simulations. We see that there are notable differences between the TNG, SIMBA, and DMO version. At group and cluster mass-scales, TNG and SIMBA both have a lower concentration than the DMO result, but this difference is subdued when looking toward the most massive objects, and this behavior is consistent with results from \citetalias{Anbajagane2022BaryImprint} (see their Figure 3). It is also promising to see that both SIMBA and TNG agree at the highest masses. This implies the fiducial model of the feedback prescriptions in TNG and SIMBA leads to broadly similar effects on $\cvir$. At smaller masses, $\Mvir < 10^{12} \msol/h$, the scaling relations have notable differences. 

In all three cases, our \textsc{Kllr} model has captured the overall dependence of $\cvir$ on simulation parameters in each version, as shown by the agreement between the solid and dashed lines. This is \textit{not} a trivial statement; even though both dashed lines and solid lines ultimately derive from using \textsc{Kllr}, they are coming from completely different sets of simulations. The solid lines are a model built using 1000 LH simulations with varying cosmological and astrophysical parameters, whereas the dashed lines are 27 CV sims at a single fiducial cosmology. Thus the agreement in Figure \ref{fig:TNG_SIMBA_DMO_CV} is one validation of our methodology.

\subsection{Astrophysics and cosmology dependence in \textsc{IllustrisTNG}}\label{sec:Astrophysics}

\begin{figure*}
    \centering
    \includegraphics[width = 2\columnwidth]{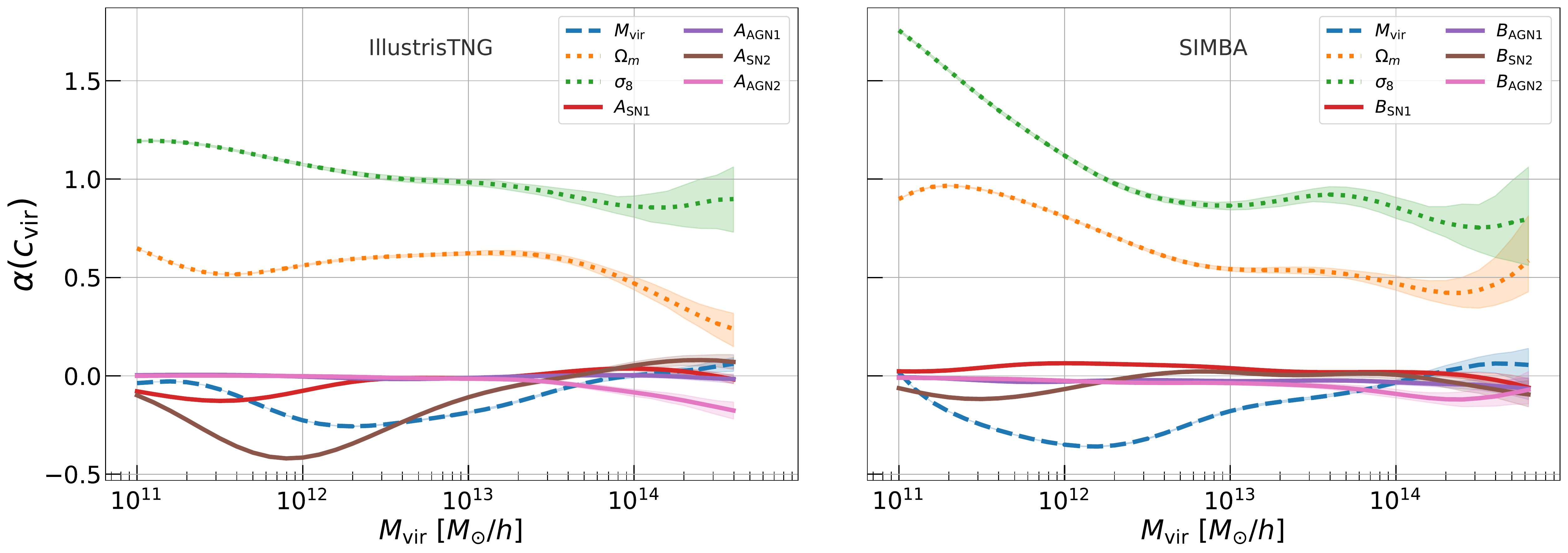}
    \caption{The $z = 0$ \textsc{Kllr} slopes of concentration $\cvir$ with halo mass $\Mvir$ and the 6 astrophysical and cosmological parameters. The bands show the 68\% confidence interval. The left (right) panel shows results from the TNG (SIMBA) sample. In all cases, the slopes have strongly nonlinear behaviors with halo mass and are mostly non-monotonic. The cosmological parameters $\Omega_{\rm m}$ and $\sigma_{\rm 8}$ have the strongest influence on halo concentrations, and are followed by the astrophysical parameters $A_{\rm SN1}$ and $A_{\rm SN2}$. The AGN parameters generally have a non-zero but low influence on $\cvir$.}
    \label{fig:TNG_SIMBA_slopes}
\end{figure*}

Figure \ref{fig:TNG_SIMBA_slopes} shows the key component of our model --- the slopes of $\cvir$ with $\Mvir$ and the 6 simulation parameters, for both TNG (left) and SIMBA (right). 
We first provide some intuition for how to interpret the slopes. Looking at the expression for our \textsc{Kllr}-based model in equation \eqref{eqn:Kllr}, we see there is an explicit mass-dependence for the slopes. In the limit that there is no such dependence, i.e. the slopes are independent of mass, they will be flat, horizontal lines in Figure \ref{fig:TNG_SIMBA_slopes}, and indicate that a simple power-law is adequate to represent the relationship between the two variables connected by the given slope parameter. A slope that changes linearly in Figure \ref{fig:TNG_SIMBA_slopes} shows preference for a linear mass term in the power-law index, \eg $X^{\alpha_X + \beta_X\Mvir}$, where $X$ is any of the 7 independent variables in the regression (halo mass or the 6 simulation parameters).

Going back to Figure \ref{fig:TNG_SIMBA_slopes}, we find strong mass-dependent features in both the TNG and SIMBA models, an expectation that was motivated by the existing works on baryon imprints that were described earlier. We see that the slope of $\cvir$ and $\Mvir$ has a clear, non-monotonic relation with a minimum roughly around the Milky Way (MW) mass-scale, though the exact location appears to vary by a factor of 2-3 in mass. This non-monotonic behavior was expected given the form of the scaling relations in Figure \ref{fig:TNG_SIMBA_DMO_CV}. The slopes for both TNG and SIMBA also have the same form of approaching zero at low and high mass-limits and having a distinct minima at MW mass-scales. It has been identified that this form arises in TNG due to gas cooling and thermal feedback competing to dominate the halo energetics \citepalias{Anbajagane2022BaryImprint}, with some indication from just group and cluster-scales that other hydrodynamic simulations show similar features \citep{Anbajagane2022GalVelBias}, and we now explicitly confirm the same qualitative behavior in SIMBA across three and a half decades in halo mass.

Moving on to the astrophysics dependence, we focus on the TNG model first. We find that the SN parameters have the strongest astrophysical impact on the halo concentration, as evidenced by their large slope amplitudes compared to the AGN slopes, and also have the most mass-dependent effect given their slopes have clear non-monotonic features (including zero-crossings) with halo mass. The SN wind speed parameter, $A_{\rm SN2}$, is the most impactful, particularly at the MW mass-scale. Following \citetalias{Anbajagane2022BaryImprint} (see their Figure 7), our expectation is for SN feedback to matter more at masses smaller than this MW-scale, as the MW mass-scale is where gas cooling and thermal (quasar) feedback become comparable; stellar feedback is subdominant. 

Thus, it may seem unexpected that the maximum impact of SN winds is at this mass scale where SN feedback is subdominant and gas cooling is dominant. However, our assumption underlying this statement is that the impact of SN is solely in how their feedback blows out the matter component and makes the halo diffuse. SN, and their winds in particular, are also a necessity for dispersing heavier elements into the halo gas component and increasing the metallicity through the halo. Increasing metallicity enhances gas cooling, which in turn increases the concentration through adiabatic contraction. For example, \citet[see their Equation 15]{Strickland2009SNEqns} show that the gas cooling rate is proportional to the mass loading factor. In the case of TNG, increasing $A_{\rm SN2}$ (with all other parameters fixed) causes the wind speed to increase, and the mass loading factor to drop. It is therefore possible for the SN model to have a noticeable baryonic imprint in halos over a wider mass-range due to its additional impact on gas cooling. The full set of processes that cause this specific mass-dependence is unclear and is beyond the scope of this work.

The other SN parameter, $A_{\rm SN1}$ also has a decent impact on $\cvir$, albeit at lower masses. This is the more ``traditional'' result of SN blowing material out the halo potential, and is reasonable behavior in CAMELS as $A_{\rm SN1}$ controls the energy output of SN and thus directly impacts the strength of these explosive events. While there appears to be a mild positive slope/correlation between $A_{\rm SN1}$ and $\cvir$ at the highest masses, the statistical significance of this slope deviating from 0 is $\lesssim 2\sigma$, and so we ignore this feature.

Next, both AGN parameters -- $A_{\rm AGN1}$ and $A_{\rm AGN2}$ --- have a weak impact on $\cvir$. In particular, $A_{\rm AGN1}$, which controls the rate at which the kinetic feedback mode accumulates energy before expelling it as a jet, has no visible impact on this relation. This is once again not particularly surprising as results in \citetalias{Anbajagane2022BaryImprint} (see Figure 7) show that in TNG the kinetic (radio) feedback is at best comparable with other galaxy formation energetics. Thus, we expect weak to no dependence here. Our slope in this case is consistent with 0 at $\lesssim 1\sigma$.

The second parameter, $A_{\rm AGN2}$, changes the ``burstiness'' of the kinetic feedback and does have an impact on $\cvir$. While this may seem counter-intuitive, since both AGN parameters impact the same kinetic feedback mode and $A_{\rm AGN1}$ clearly has no impact on $\cvir$, there are some key differences. Recall our previous discussion in Section \ref{sec:TNGDescription} that the kinetic feedback in TNG works by accumulating energy until a set energy scale is reached and then by expelling it as kinetic energy in the form of momentum kicks. This is what creates the ``burstiness'' in the kinetic jets. $A_{\rm AGN1}$ changes how quickly/slowly the SMBH achieves this energy threshold to then create a jet. Whereas $A_{\rm AGN2}$ changes the exact energy scale of the jet, which also sets the minimum kinetic energy injected into the halo. Although kinetic feedback is subdominant in the fiducial TNG model, varying $A_{\rm AGN2}$ changes the strength of individual feedback events, which can alter this behavior. As a toy example, 100 temporally spaced out AGN events of $10^{50} \rm ergs$ can have a subdominant/incremental impact compared to a single event of $10^{52} \rm ergs$. The parameter $A_{\rm AGN2}$ sets where we are on this scale of ``many, weaker events'' and ``fewer, stronger events''. We also stress that the dependence of $\cvir$ on $A_{\rm AGN2}$ is still mild for the mass ranges explored in this work, though it is possible it is stronger at the most massive halo scales $\Mvir > 10^{15} \msol/h$ that we were unable to probe here.

Finally, we come to the slopes of the different cosmology parameters $\Omega_{\rm m}$ and $\sigma_8$, which have the strongest impact, i.e. the highest slopes, on $\cvir$. The slopes do have some mass-dependence, especially for $\Omega_{\rm m}$ whose slope shows a ``wiggle''-type feature. When running a full hydrodynamic simulation, we expect the cosmology dependence is ``convolved'' with the suite of baryon evolution effects, and given galaxy formation physics has many different mass/energy/length scales, we will expect the cosmology dependence to also become strongly scale-dependent, with potentially non-monotonic features. This is exactly what we see. However, to zeroth-order the slopes' change with mass is consistent with that seen in DMO simulations, and we will discuss this further in Section \ref{sec:TNGvsDMO}.

Looking at the slopes between $\cvir$ and the cosmology parameters, one may wonder why the astrophysical dependence is deemed important given the amplitude of slopes between $\cvir$ and astrophysical parameters can be a factor of 2-3 lower than that between $\cvir$ and the cosmological parameters. However, we know the value of cosmological parameters very well. Measurements of $\sigma_8$ and $\Omega_{\rm m}$ have resulted in 1-2\% constraints \citep[\eg][]{Planck2015CosmoParams, Asgari2021KidsWL, DES2022Y3}. This causes a shift of 1\% (0.5\%) in $\cvir$ due to any shifts in $\sigma_8$ ($\Omega_{\rm m}$) resulting from the uncertainty. On the other hand, we do not have a good handle for priors on the astrophysical parameters. Going by the range of values chosen in \textsc{Camels}, they can vary by $150-250\%$, and so the concentration can change by up to $100\%$. This is just a back-of-the-envelope estimate meant to motivate why the astrophysical impacts are still important (and in fact, likely dominant) here, even though the slopes with cosmological parameters have larger amplitudes.

\subsection{Comparison of astrophysics and cosmology dependence in \textsc{IllustrisTNG} and \textsc{Simba}}\label{sec:TNGvsSIMBA}

As was described before, the TNG and SIMBA galaxy formation models are different, and the \textsc{Camels} simulation parameters modify different equations within each model. Thus, the connection between $\cvir$ and the simulation parameters is altered across models. We detail the differences here.

First, the slopes with $\Mvir$ have different detailed behaviors. This difference is a combination of all differences in the physics of each galaxy formation model. We do not focus on the details of these differences in this work as it is difficult to disentagle all the input effects to make causal statements. The general point we make, however, is that the mass-dependence of the slope with $\Mvir$ in both simulations is qualitatively the same --- the slope is closer to zero at both high and low masses, but has a strongly negative value at Milky Way mass scales.

Now focusing on the 6 simulation parameters, we first find that the cosmology dependence is quantitatively --- and to some extent qualitatively --- different between the TNG and SIMBA models. This is expected given our previous discussion about how these slopes with $\Omega_{\rm m}$ and $\sigma_8$ capture the effect of cosmology convolved with all the physics of a galaxy formation model. As the galaxy formation model changes, the cosmological dependence will be affected as well. In general, both the TNG and SIMBA models show the impact of cosmology is more strongly felt at lower masses, particularly in SIMBA, and we show below that this is a feature that exists in the DMO results as well (Section \ref{sec:TNGvsDMO}). The ``wiggle'' feature in the $\Omega_{\rm m}$ slope of TNG is not seen in SIMBA.

Moving onto the four astrophysical parameters, we find more differences here. Notably, the slope with $B_{\rm SN1}$ takes a positive value in SIMBA, even at small masses, whereas for $A_{\rm SN1}$ in TNG this slope was negative. This can once again be attributed to cooling --- $B_{\rm SN1}$ in SIMBA directly controls the wind mass-loading (whereas in TNG the combination of $A_{\rm SN1}$ and $A_{\rm SN2}$ controlled this), so as $B_{\rm SN1}$ is increased, the cooling rate increases \citep[\eg][see their Equation 15]{Strickland2009SNEqns}. This in turn causes a higher $\cvir$, leading to the positive slope between $\cvir$ and $B_{\rm SN1}$. Next, $B_{\rm SN2}$ alters the wind speed in SIMBA, just like $A_{\rm SN2}$ in TNG, and so the behavior is qualitatively similar in the mass-dependence as $\{A,B\}_{\rm SN2}$ in both TNG and SIMBA are consistent with zero at high mass and become more negative towards low mass. Differences in the actual amplitude of the slopes will arise from more technical differences in the implemented model and we do not focus on those here.

Moving to the AGN parameters, we find once again that the parameters do not impact the $\cvir-\Mvir$ relation much, as evidenced by slope amplitudes being close to zero. Of note is the fact that in SIMBA, the $B_{\rm AGN1}$ parameter changes the kinetic energy scale of the jets, whereas $A_{\rm AGN1}$ in TNG changes the energy accumulation rate. The $B_{\rm AGN2}$ in SIMBA controls the speed of kinetic feedback jets, which is directly analogous to the $A_{\rm AGN2}$ parameter in TNG. Thus, we see that $B_{\rm AGN2}$ is the more impactful of the two SIMBA AGN parameters, is mostly impactful at high halo mass, and works to reduce the concentration.

\subsection{Redshift dependence in \textsc{IllustrisTNG}}\label{sec:RedshiftDependence}

\begin{figure*}
    \centering
    \includegraphics[width = 2\columnwidth]{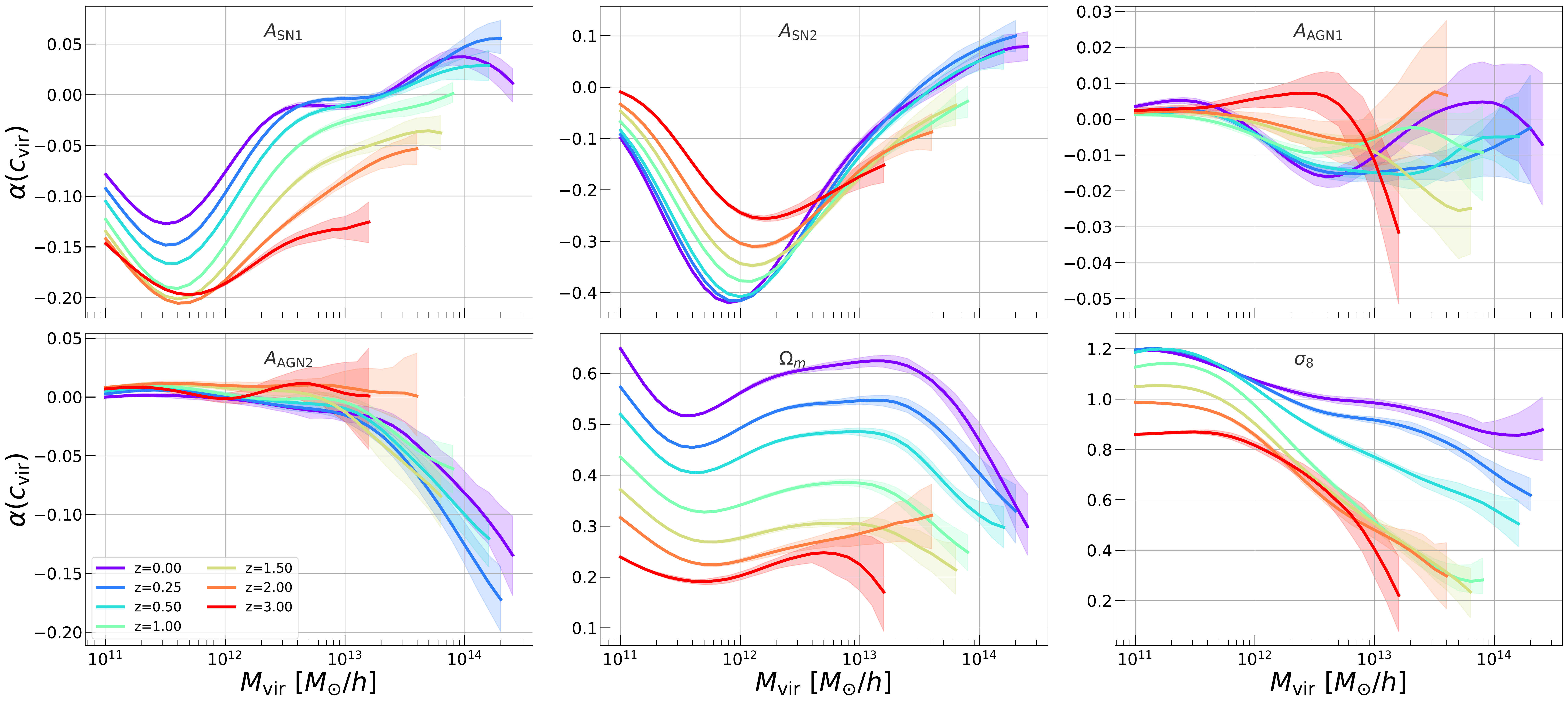}
    \caption{The \textsc{Kllr} slopes between halo concentration $c_{\rm vir}$ and the six astrophysical and cosmological parameters (denoted by text in each panel) at redshifts $0 < z < 3$. The slopes are interpolated across redshift to get the values at the exact redshift of interest. The bands show the 68\% confidence intervals. There is a clear and strong mass-dependence in all parameters, and a significant redshift evolution in most.}
    \label{fig:slopes_z_dep}
\end{figure*}

We have thus far described the behavior of the relations between halo concentration and the simulation parameters at $z = 0$. However, these relationships will not be constant over redshift, given that galaxy formation processes vary strongly across redshift \citep[see Figure 1 in ][]{Weinberger2018SMBHsIllustrisTNG}; for example, star formation efficiency is known to peak at $z = 2$ and is not constant over time \citep[\eg][Figure 3]{Behroozi2013StarFormationHistory}.

In Figure \ref{fig:slopes_z_dep}, we show the slopes between $\cvir$ and the six simulation parameters in \textsc{IllustrisTNG} as a function of redshift for the range $0 < z < 3$. The results for \textsc{Simba} are extracted in Appendix \ref{appx:SIMBA}. In practice, to get these results, we use the 34 \textsc{Kllr} models built at different redshifts to create an interpolator\footnote{We use the \textsc{interpn} function from the \textsc{scipy} package \citep{Virtanen2020Scipy}, and interpolate over a regular, rectangular grid of the mass and redshift input variables.} over $\log_{10}\Mvir$ and $\log_{10}(1 + z)$. This interpolator is used to get predictions at the exact mass and redshifts of interest. We use the 100 realizations of the \textsc{Kllr} model to create multiple interpolators, which then give us multiple predictions and quantify uncertainties in these predictions. 

In general, the shape of the slopes' mass-dependence seen at $z = 0$ is preserved across a wide range in redshift, though the exact values of the slopes differ significantly. We now examine the specific behavior in each slope.

\textbf{In $\boldsymbol{A_{\rm SN1}}$}, the slopes become more negative with time across all masses. This is reasonable as star formation becomes much more efficient around $z \sim 2$, so the energy scale of SN feedback is more relevant/impactful at those redshifts. We see a characteristic dip at $\Mvir \approx 10^{11.5}\msol/h$ that is mostly constant over redshift. The slopes also have a plateauing phase around $10^{13} \msol/h$ which is seen at all redshifts, but appears to happen at higher masses at higher redshifts. Multiple redshifts also show the characteristic rise after the plateauing that gets stronger towards galaxy cluster mass-scales.

\textbf{In $\boldsymbol{A_{\rm SN2}}$}, there appear to be two different mass regimes. At low masses, the slopes show the same shape for the mass-dependence, but get more shallow with redshift. At higher masses, the mass-dependence is nearly constant with redshift, and appears to tend towards positive values at the highest masses. These two regimes are split at around the same mass of $10^{12.8} \msol/h$ for all redshifts. For the low-mass dependence, the exact minima of the slopes shifts in mass as well; at higher redshift, it occurs at higher mass.

\textbf{In $\boldsymbol{A_{\rm AGN1}}$ and $\boldsymbol{A_{\rm AGN2}}$}, the slopes are mostly consistent with zero across all mass ranges and redshifts. The strongest impact of $A_{\rm AGN1}$ is around $\Mvir \approx 10^{12.5} \msol/h$, where the slope deviates the most from zero; but this is only at $z = 0$, as this slope returns back to 0 at $z = 3$. For $A_{\rm AGN2}$, the slopes above $\Mvir > 10^{13}\msol/h$ move away from zero and become negative. At the highest masses, the level of impact, shown by the amplitude of the slope, is comparable to the maximum impact $A_{\rm SN1}$ has. The lack of redshift dependence is consistent with how, at the highest halo mass, kinetic AGN feedback in TNG is quite constant across redshift \citep[Figure 1]{Weinberger2018SMBHsIllustrisTNG}.

\textbf{In $\boldsymbol{\Omega_{\rm m}}$}, we see the strongest redshift dependence. Once again, the shape of the mass-dependence is mostly constant, but the amplitude varies significantly, where at higher redshift there is a lower impact on $\cvir$. There is a peak in the slopes at $\Mvir = 10^{11} \msun/h$ that is prominent at low redshifts and diminishes at higher redshifts. The slope also seems to tend towards $\alpha \sim 0.25$ at the highest masses, and this is true for all redshifts where we can probe the mass range $\Mvir > 10^{13.5} \msol/h$. There is also a local minima at $\Mvir = 10^{11.5} \hinv \msol$, which is interestingly the same mass-scale where we found a minimum in the $A_{\rm SN1}$ slopes. For this work, we do not pursue any causal link connecting the two features.

\textbf{Finally, in $\boldsymbol{\sigma_8}$}, we have differences in the shape of the mass-dependence across redshift. At low masses, there is a clear plateau in the slopes at all redshifts, with the plateau value decreasing with redshift. At higher masses, the slope decreases more steeply with mass as we increase redshift, though the increase in steepness reaches an asymptotic value after $z = 1$, especially for $\Mvir > 10^{12.5} \msol/h$.

In general, Figure \ref{fig:slopes_z_dep} shows a rich phenomenology for how different galaxy formation processes impact $\cvir$ as a function of redshift. It also provides an example of the scale-dependence of galaxy formation discussed previously. We have thus far seen this scale-dependence as a non-monotonic mass-dependence of the slopes, but we now explicitly see this in the redshift-dependence as well. Notably, there is no simple redshift-based rescaling of $\cvir$ or the slopes $\alpha_X$ that can give us the right redshift-dependence for all six parameters at once. Each parameter must be treated differently, given it is addressing different galaxy formation processes which are all mass- and redshift-dependent in their own unique way.

While we have discussed only the TNG here, this statement on the mass- and redshift-dependence of the impact of galaxy formation on halos is also true in SIMBA. We show and briefly discuss those results in Appendix \ref{appx:SIMBA}.

\subsection{Comparison of redshift evolution in \textsc{IllustrisTNG} and DMO}\label{sec:TNGvsDMO}

\begin{figure*}
    \centering
    \includegraphics[width = 2\columnwidth]{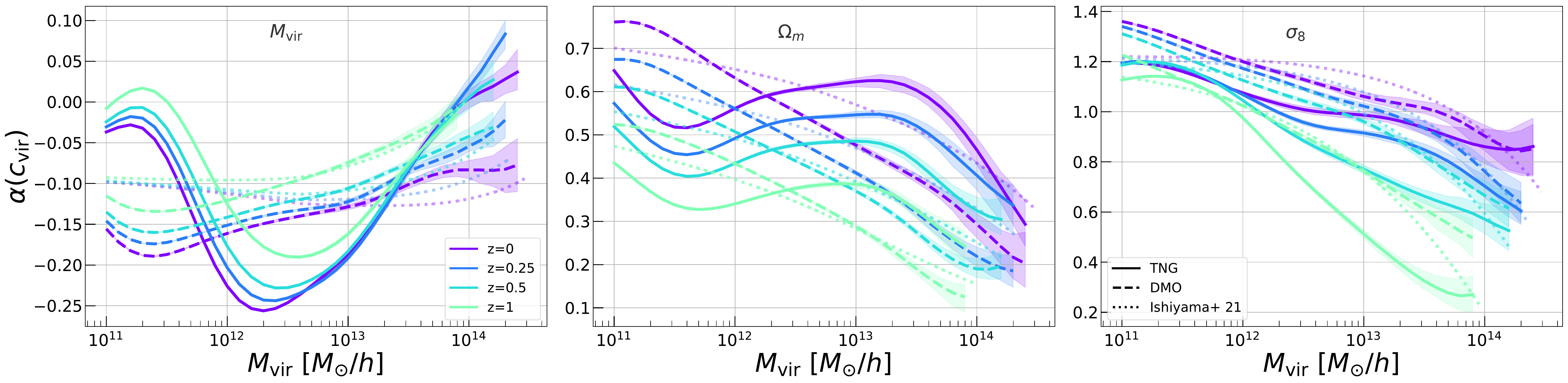}
    \caption{The \textsc{Kllr} slopes of halo concentration $\cvir$ with halo mass $\Mvir$, $\Omega_{\rm m}$ and $\sigma_8$ at $0 < z < 1$. The format is the same as Figure \ref{fig:slopes_z_dep}. We compare the TNG slopes (solid) with the slopes from the DMO simulations (dashed) and predictions from the model of \citet{Ishiyama2020UchuuConcentration} using the \textsc{Colossus} package \citep{Diemer2018COLOSSUS}. The mass-dependence of the slopes in the DMO model and \textsc{Colossis} have similar trends with mass, with the slopes decreasing towards larger mass-scales. The deviations between the TNG and DMO runs, however, are qualitatively different for $\Mvir$ and $\Omega_{\rm m}$, and less so for $\sigma_8$.}
    \label{fig:TNG_DMO_Colossus}
\end{figure*}

Finally, we discuss the differences between the slopes in TNG and those in the DMO simulations of \textsc{Camels}. Figure \ref{fig:TNG_DMO_Colossus} shows the slopes for TNG and DMO, in addition to theoretical predictions using the model of \citet{Ishiyama2020UchuuConcentration} as implemented in \textsc{Colossus} \citep{Diemer2018COLOSSUS}.\footnote{We compute the slopes for the \citet{Ishiyama2020UchuuConcentration} model by taking numerical derivatives, $\Delta\ln \cvir/\Delta\ln p = (\ln \cvir(p^+, ...) - \ln \cvir(p^-, ...))/\Delta \ln p$, where $p$ is one of the parameters of interest.} For slopes of $\cvir$ with $\Mvir$ and $\Omega_{\rm m}$, the TNG results shows more complexity than either the DMO or Ishiyama model. The slopes with $\Mvir$ have a minimum near $10^{12} \msun/h$ and slopes with $\Omega_{\rm m}$ have a maxima near $\rm 10^{13.5} \msun/h$. Both features are not presented in the DMO or Ishiyama models, which always show monotonic trends for the dependence of the slope on halo mass. For the slopes of $\sigma_{8}$, the TNG model has qualitatively similar trends as the DMO and Ishiyama model in that it uniformly decreases with mass. In each of the three slopes (panels), the redshift dependence is similar among the three models shown. At higher redshift, the slopes of $\Mvir$ increase and those of $\Omega_{\rm m}$ and $\sigma_{\rm 8}$ decrease.

Our findings confirm that the slopes with mass and cosmological parameters is to zeroth order given by the DMO behavior. However, there are notable deviations --- with the deviations' amplitudes varying across $\Mvir$, $\Omega_{\rm m}$ and $\sigma_8$ --- for the slopes of $\cvir$ with all three quantities and these are caused by galaxy formation processes.

\section{Results II: Model applications \& Parameter Constraints}\label{sec:ModelApplications}

Having now discussed the physics extracted from examining our \textsc{Kllr}-based model, we now move towards applications of the model. We first start by describing some potential applications in \S \ref{sec:Model}. Then we show two specific applications --- a simulation ``translation'' to match TNG and SIMBA in \S \ref{sec:SimTranslation} and an observational constraint in \S \ref{sec:ObsConstraint}.

\subsection{Model features and applications}\label{sec:Model}

\begin{figure*}
    \centering
    \includegraphics[width = 2\columnwidth]{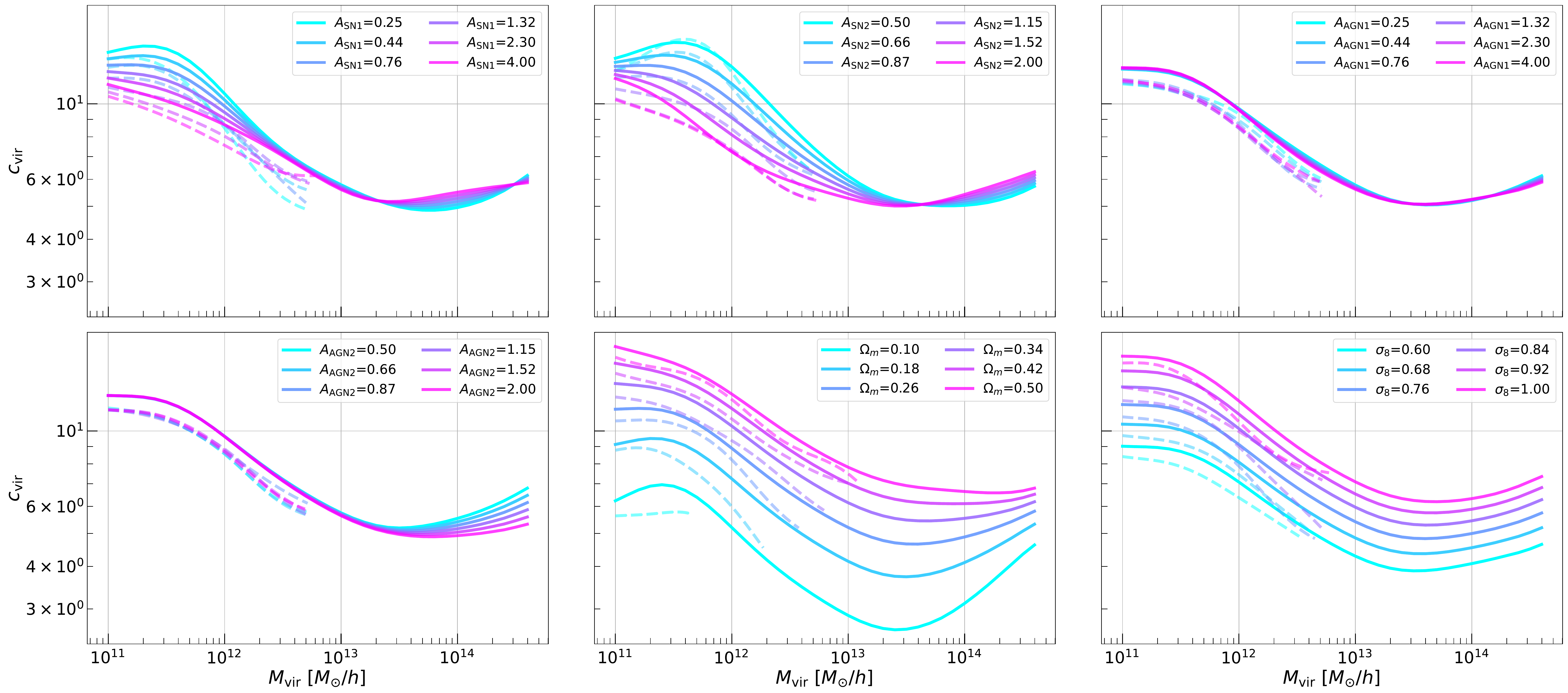}
    \caption{The $z = 0$ TNG \textsc{Kllr} scaling relations for $\cvir$ with various simulation parameters (shown in the legends) and the color tones show these relations as we vary the parameters. The solid lines are the prediction from our model, made using the 1000 LH simulations, and the dashed lines are the \textsc{Kllr} result from the 1P simulations. We find that the model generally picks up the same trends as those found in the simulation results. Note that we do not expect precise agreement between the model and the 1P simulations here given the 1P simulations all share the exact same initial conditions and so cosmic variance (which is particularly impactful here given small boxsize) can shift a single simulation's result high/low. The relative difference between different colors of the same linestyle is similar across the  two linsetyles.}
    \label{fig:TNG_1P}
\end{figure*}

By extracting the slopes of the halo concentration with mass and the six simulation parameters, we have built a non-linear (but locally linear) model for the $\cvir-\Mvir$ relation as a function of astrophysical and cosmological parameters. While \textsc{Kllr} is a non-parameteric non-linear model (i.e. it apriori knows nothing about the scaling relations) its connection to linear regression, through the common language of intercepts and slopes, allows for an easy and familiar physical interpretation of the model results. 
In Figure \ref{fig:TNG_1P} we show that this model correctly captures the dependence of the nonlinear $\cvir-\Mvir$ relation on cosmological and astrophysical parameters, as determined by the TNG galaxy formation model. In all panels, the mass-dependent evolution of $\cvir$ as a parameter is varied is similar across our model (solid lines) and the 1P simulations (dashed lines). There are some offsets, but we are primarily focused on the relative trends between different colors of a given linestyle rather than offsets between linestyles. This is because the 1P simulations are a single run with the same initial conditions rather than the ``true mean'' obtained by averaging over many realizations.

As discussed before, our model consists of the intercepts and slopes needed for equation \eqref{eqn:Kllr} and this is done for 34 redshifts and for both galaxy formation models. These outputs can be used to construct an interpolator that efficiently make predictions at the exact mass and redshift needed for any analysis; we have performed such an exercise for in the above sections and will further make use of this tool in the sections below.

In many aspects of cosmology and astrophysics, we use the $\cvir-\Mvir$ relation to forward model different observables. Some well-known examples on the structure formation side are the small-scale matter power spectra \citep[\eg][]{Zacharegkas2021GGLensingDES}, and cross-correlations between various fields \citep[\eg][]{Pandey2019GalaxytSZ, Pandey2021DESxACT, Gatti2021DESxACT}. It is also used extensively in analytic and numerical studies of satellite halo dynamics \citep[\eg][]{VanDenBosch2018Subhalo, Shen2022AxionMini}. All of these works use relations calibrated from DMO simulation suites \citep[\eg][]{Bhattacharya2013Concentrations, Diemer2015Concentration, Diemer2019concentrations, Ishiyama2020UchuuConcentration}. The relations calibrated in our work, however, extend these models in a complementary way to include baryon imprints from galaxy formation processes. In doing so, we provide a simple, yet powerful, method for incorporating astrophysical effects into any halo model-based analysis. Of course, one could directly measure the relevant observable, such as the matter power spectrum or the gas-matter cross power spectrum, in \textsc{Camels} and build a calibration framework for that observable specifically. However, calibrating the halo model (i.e. the halo profiles) is inherently more intuitive given baryonic effects occur \textit{within} halos, and it is also more flexible in that it allows \textit{any} halo model-based prediction for a matter density field observable to incorporate these baryonic effects.

In general, our model captures the simulation results to the $10\%$ accuracy level. This is estimated by taking each LH sim and computing the residuals between the sim's \textsc{Kllr} fit --- which is taken as the ground truth --- and our model's prediction for that LH sim. We then take the residuals from the ensemble of 1000 available LH simulations and compute the 68\% interval for  these residuals as a function of mass, which leads us to the estimate of 10\% accuracy. This number is best representative of the accuracy at low masses ($\Mvir < 10^{13} \msol/h$). At higher masses, we cannot reliably compute the accuracy as the \textsc{Kllr} fit used a ground truth has a cosmic variance component which dominates the model accuracy and thus would dominate the residuals. We have also tested the accuracy improvement gained by including higher, quadratic terms, such as $A_{\rm X}^2$ and $A_{\rm X}A_{\rm Y}$, and found improvements of only 1-2\%. Thus, we optimize for simplicity in the model and do not include these higher-order terms in the local linear regression described in equation \eqref{eqn:Kllr}.

One subtle limitation of our model is our mass range; we have an upper bound of $\Mvir \approx 10^{14.5} \msol/h$ at $z = 0$, and this decreases with redshift. Some observables, such as the cross power spectra between gas pressure and total matter fields, have a significant contribution from halos above this mass range \citep[see Figure 2 of][for an example]{Pandey2021DESxACT}. Though, the exact mass range that is relevant depends on the redshift/length-scale being studied. Other observables such as the lensing power spectra or total matter power spectra have dominant contributions from lower mass halos that are well-represented in our sample and have only diminished contributions from halos at the upper edge of our sample's mass range. The lower bound of $\Mvir = 10^{11} \msol/h$ in this work, set by resolution constraints, is adequate for most astrophysical and all large-scale structure modeling. Our redshift range, with sufficient statistics up to $z = 3$ and a strict cut at $z = 6$, is adequate for almost all studies of late-time structure formation.

\subsection{Simulation translation}\label{sec:SimTranslation}

\begin{figure*}
    \centering
    \includegraphics[width = \columnwidth]{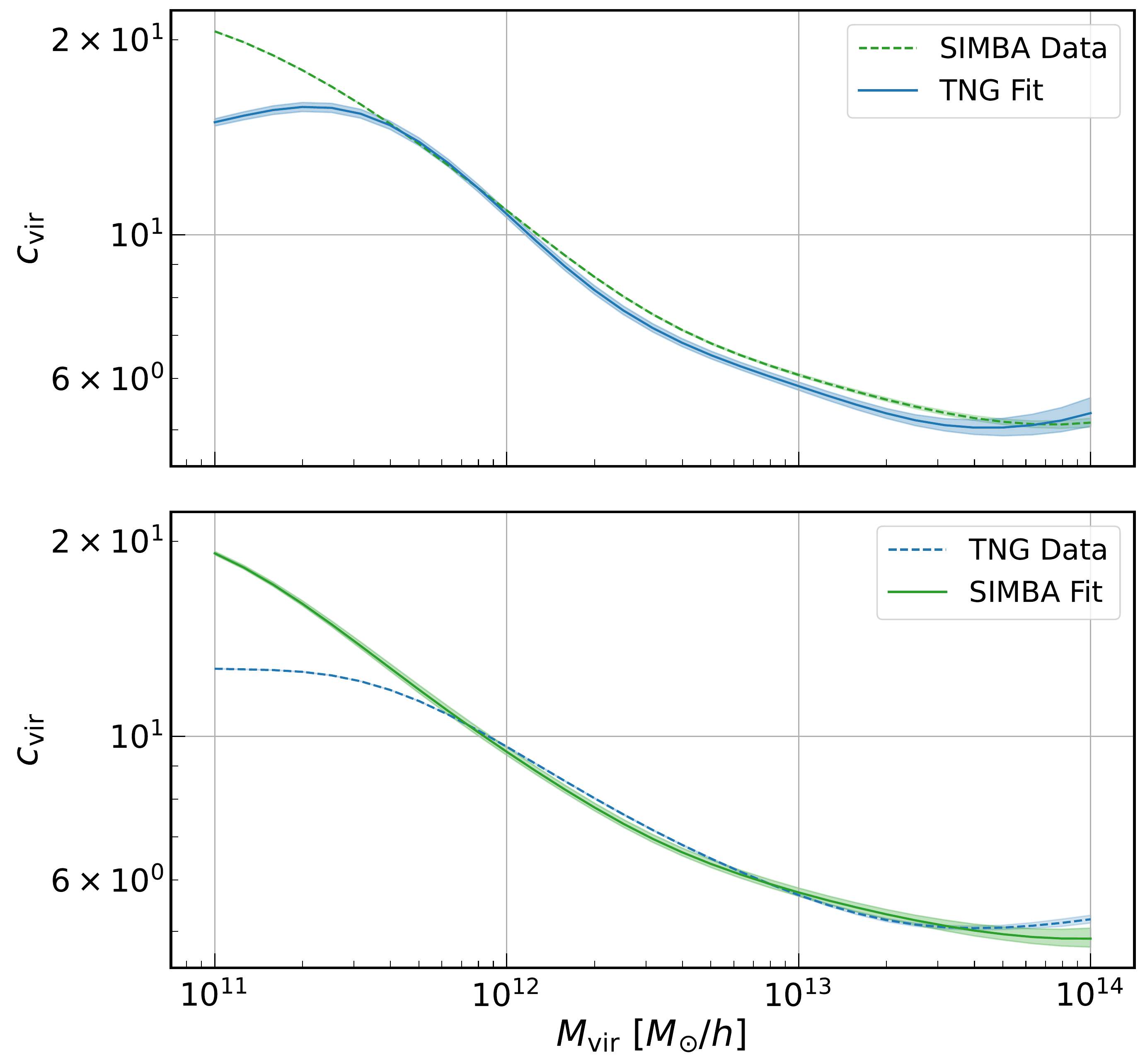}
    \hspace{15pt}
    \includegraphics[width = 0.93\columnwidth]{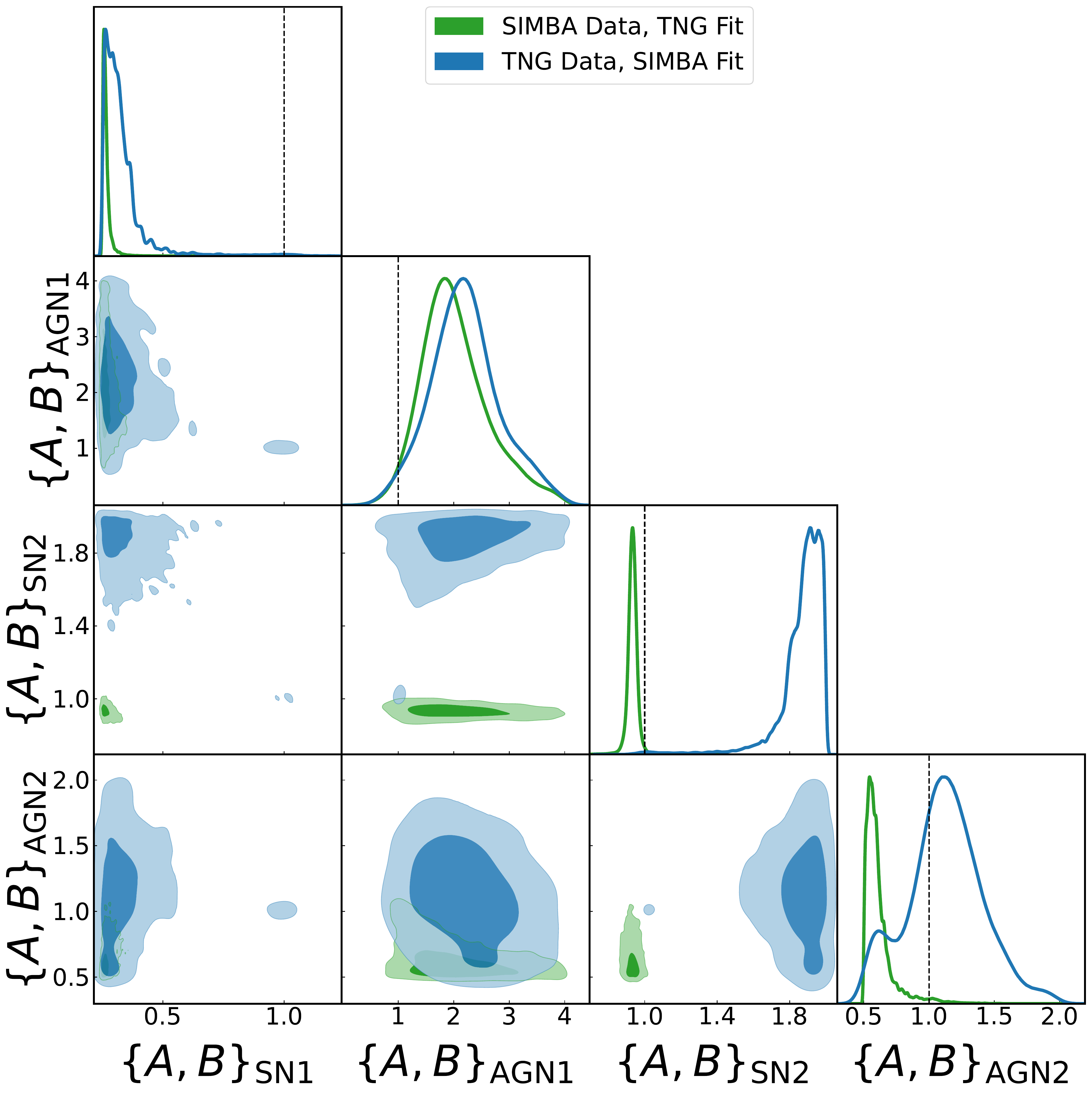}
    \caption{Results from doing a ``translation'' between simulations using the \textsc{Kllr} model. \textbf{Left:} The fiducial $\cvir-\Mvir$ relation for each model \textbf{A} (denoted as ``Data'') and the accompanying fit using the alternative model \textbf{B} (denoated as ``Fit''). Results are obtained from running an MCMC. The uncertainties shown in the ``Data'' come from the boostrap realizations of the interpolated \textsc{Kllr} predictions, whereas for the fit they come from the MCMC posteriors. \textbf{Right:} The posterior distributions for the 4 astrophysical parameters of \textsc{Camels} in each galaxy formation model. \textit{Note that the meaning of the parameters is different between TNG and SIMBA}. These plots show that our model complexity retains enough flexibility to have the results of one galaxy formation model match that of another for a large part of the mass range.}
    \label{fig:SimTranslation}
\end{figure*}

An interesting application of our model is to take the predictions from galaxy formation model \textbf{A}, obtained using the fiducial parameter values, and see how the astrophysical parameters in a model \textbf{B} must deviate from their fiducial setting so the prediction of model \textbf{B} matches that of model \textbf{A}. This is in effect probing an effective ``translation'' between model \textbf{B} and model \textbf{A}.

In Figure \ref{fig:SimTranslation} we show results from this procedure. In specific, we run a Bayesian analysis, using Markov chain Monte Carlo (MCMC), to determine the astrophysical parameters that make TNG and SIMBA match each other. We fix the cosmology parameters to $\Omega_{\rm m} = 0.3$ and $\sigma_8 = 0.8$, which are the fiducial values. The astrophysical parameters are given a prior of $0.25 < \{A, B\}_{\rm SN1}, \{A, B\}_{\rm AGN1} < 4$ and $0.5 < \{A, B\}_{\rm SN2}, \{A, B\}_{\rm AGN2} < 2$, which come from the range of values used in the LH sims in \textsc{Camels}.

The MCMC is run using the \textsc{emcee} package \citep{Foreman-Mackey2013emcee}. For the covariance matrix of the ``data'' (which is the prediction of model \textbf{A}), we use a modified matrix. In general, one would get the covariance matrix using the 100 bootstraps, and that is perfectly valid. Note that the variance on the mean $\cvir$ increases significantly as a function of $\Mvir$ given the halo counts drops exponentially with mass. This means the MCMC is incentivized to fit the $\cvir -\Mvir$ relation at the smallest mass-scales, where even moderate improvements can greatly increase the chi-squared.

Our goal, on the other hand, is to create a toy scenario where we first assume the mean prediction of model \textbf{A} is the ``true mean'' (i.e. we know the right mean $\cvir$ at every $\Mvir$ with no measurement uncertainty) and then ask how model \textbf{B} can replicate that ``true mean''. In this context, it is undesirable to have the MCMC fit just the lowest masses. Thus, we ``normalize'' the covariance matrix so the concentration at all masses are prioritized equally in the fit. We first start by computing the correlation matrix $r_{ij}$ and then convert it to a covariance matrix assuming a constant scatter of $1\%$, in the appropriate $\rm dex$ units. By forcing the variance of the measurements to be mass-independent, we make the MCMC procedure fit all masses equally.\footnote{This can also be viewed as regularizing the optimization function, which is simple least squares regression, to include some mass-dependence where measurements at lower masses are downweighted to offset their high precision relative to those at higher masses.} 
Note that with this procedure we still keep the correlation structure of the covariance matrix, i.e. we do not ignore any off-diagonal terms. We have also checked that varying the value of the $1\%$ constant scatter does not affect our final results

We perform this analysis in two settings, each time swapping which of TNG and SIMBA serve as model \textbf{A} and model \textbf{B}. Both results are shown in Figure \ref{fig:SimTranslation}. 
Qualitatively we find that for TNG to match SIMBA, we must reduce the SN energy scale ($A_{\rm SN1}$) and reduce the AGN jet speed and burstiness ($A_{\rm AGN2}$). The other parameters are within $< 2\sigma$ of their fiducial values. To match SIMBA with TNG, we need to reduce the mass loading factor ($B_{\rm SN1}$) and increase the supernova wind speed ($B_{\rm SN2}$). The AGN parameters are also within $< 2\sigma$ of their fiducial values

In general the preference of SN parameters in both cases is expected. A look at the dashed lines on left panels of Figure \ref{fig:SimTranslation}, which are the fiducial relations from TNG and SIMBA, shows that TNG has a distinct overturn at low masses which is not present in SIMBA. In TNG this feature is caused by the interplay between cooling increasing concentration at MW-scales and SN feedback reducing it at lower masses \citepalias{Anbajagane2022BaryImprint}. A similar feature is not present at the same mass-scale in SIMBA, likely due to a weaker SN feedback or more dominant cooling, which will keep increasing the concentration without any overturn. Notably, the mass loading factor in SIMBA is roughly an order of magnitude higher at the smallest masses when compared to the corresponding values in TNG \citep[see their Figure 1]{Dave2019SIMBA}, and more mass loading increases the metal enrichment of the halo gas, leading to more cooling and efficient adiabatic contraction. So to match the predictions of the two simulations, one must either modify TNG to reduce SN power (reduce $A_{\rm SN1}$), or change SIMBA to increase SN power (increase $B_{\rm SN2}$) and/or reduce gas cooling (decrease $B_{\rm SN1}$).

For TNG to match SIMBA we see that $A_{\rm AGN2}$ must take values below the fiducial setting of $A_{\rm AGN2} = 1$. Given our current results, we are unable to form a more informative picture beyond realizing that TNG prefers to have weaker but more frequent kinetic jet feedback events. It is also worth noting that simulations run with versions of the SIMBA model modified for galaxy cluster simulations \citep{Cui2022SIMBA} have been found to heat the gas considerably more compared to the models of TNG and other simulations \citep{Lee2022rSZ}. This suggests the AGN feedback in SIMBA has a \textit{stronger} impact on its environment than in TNG. This is also supported by studies of the baryon radial distribution in TNG and in the original, unmodified SIMBA model \citep*{Ayromlou2022ClosureRadius}.

In general, a key takeaway is we are able to make one model match the other in the mass range of $\Mvir > 10^{11.5} \msol/h$ or $\Mvir > 10^{12} \msol/h$, depending on which model is the ``data''. Right at and below these mass ranges the difference in galaxy formation physics is far too large to allow us to ``translate'' between the models. Conversely, this means if we can obtain precise observational measurements of the concentration at these lower masses ($\Mvir < 10^{12} \msol/h$), we will be able to better determine which of the TNG and SIMBA models more closely match the real Universe. Note that while the TNG and SIMBA fiducial relations do have mass ranges where they are generally in good agreement, this happens above $\Mvir > 10^{12.5} \msol/h$ but still with some visible deviations (Figure \ref{fig:TNG_SIMBA_DMO_CV}). By varying the astrophysical parameters input to the model, we get the agreement to improve over all mass-scales and also to stretch down in mass by an order of magnitude.

In addition, while we are varying four highly-relevant astrophysical parameters, there are additional processes with more nuanced effects that we do not vary given \textsc{Camels} does not vary the equations. So we apriori do not expect to be able to precisely convert results between simulations given the modest degrees of freedom in our model. For example, \citetalias{Anbajagane2022BaryImprint} found that the quasar/thermal feedback mode was the most relevant process at $\Mvir \approx 10^{12} \msol/h$ scales, and also that the SMBH mass-scale setting the transition from thermal to kinetic feedback played a role in shifting the nonlinear mass-dependence of $\cvir$; both are parameters that are not varied in the TNG version of \textsc{Camels}. Finally, we note that we have only performed this ``translation'' exercise for $z = 0$. 

Note that using TNG as the model in the fitting procedure results in tighter posteriors for the parameters (i.e. blue posteriors are tighter than green in Figure \ref{fig:SimTranslation}), and this is simply because the slopes between $\cvir$ and the parameters in TNG are generally of a larger amplitude, which naturally results in tighter constraints on said parameters.

\subsection{Observational constraints}\label{sec:ObsConstraint}

While we have used our \textsc{Kllr}-based model so far to get parameter constraints ``translating'' between simulation predictions, we can also use observational measurements on the $\cvir-\Mvir$ relation to obtain constraints on the simulation parameters for a given galaxy formation model. The data we use for this analysis is described in Section \ref{sec:ObsData} and is composed of results from multiple works \citep{Mandelbaum2006SDSSCvir, Johnston2007Concentration, Mandelbaum2008ConcentrationMeasurement, Brimioulle2013CFHTCvir}.

\begin{figure}
    \centering
    \includegraphics[width = \columnwidth]{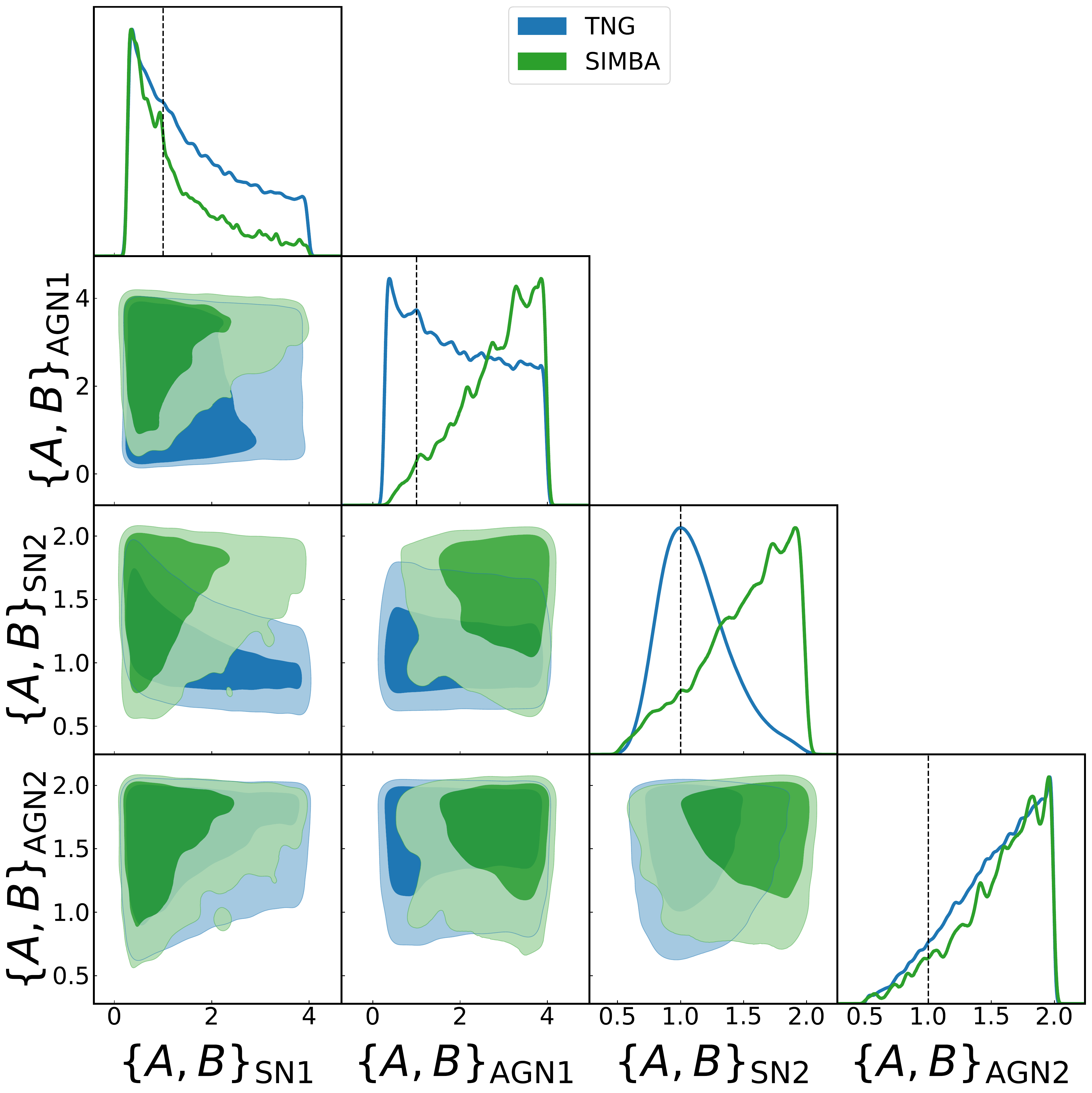}\\[20pt]
    \includegraphics[width = \columnwidth]{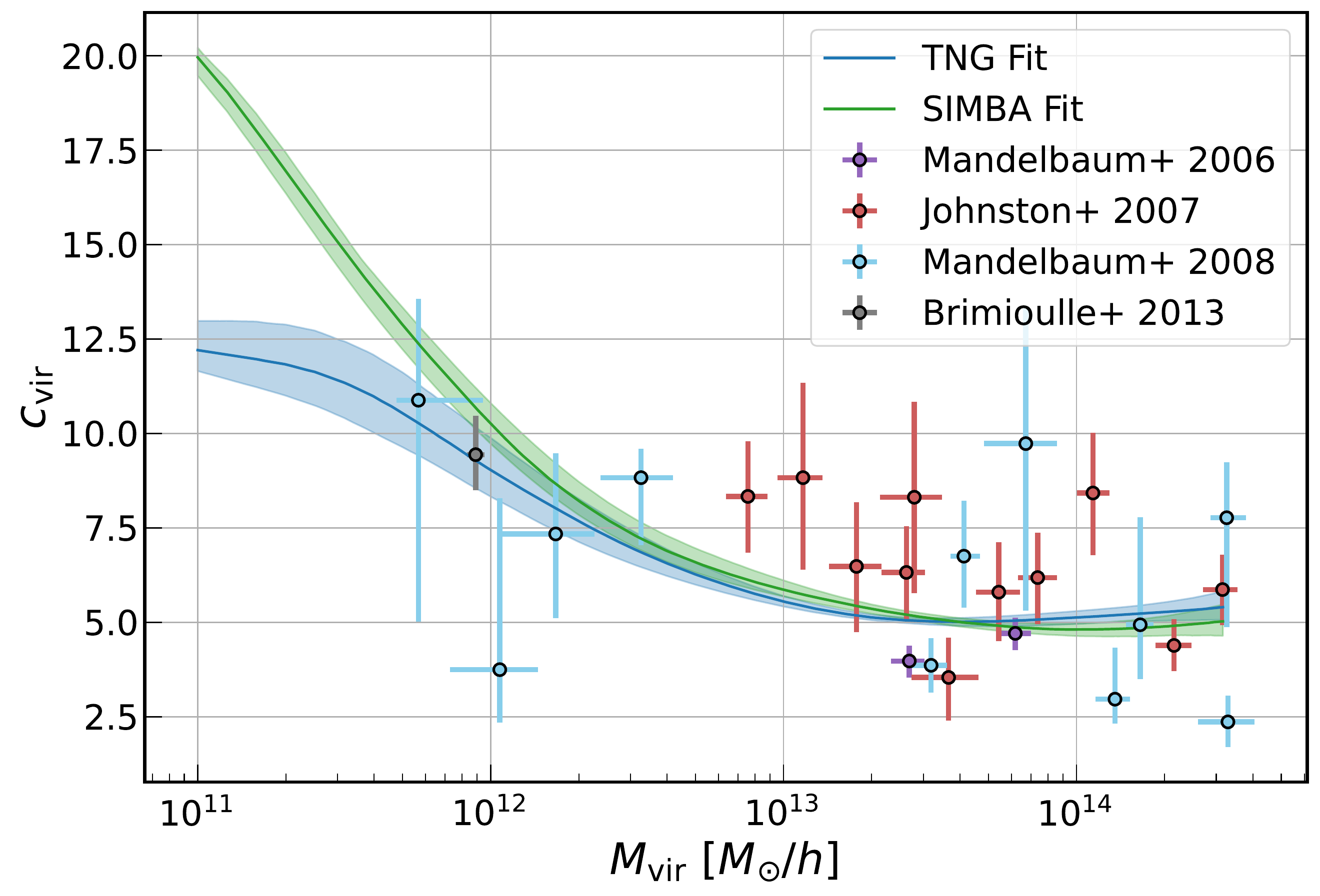}
    \caption{\textbf{Top:} The posteriors on the \textsc{Camels} astrophysical parameters from fitting the two galaxy formation models to observational data. 
    \textbf{Bottom:} The $\cvir-\Mvir$ relations for TNG and SIMBA models as determined by the MCMC fits, and the data points used in the analysis. The two models generally agree above $10^{12}\msol/h$ and start disagreeing more at lower masses where we have no data.}
    \label{fig:ObsConstraints}
\end{figure}

We analyze this data using both the TNG model and the SIMBA model, and results are shown in Figure \ref{fig:ObsConstraints}. We find, unsurprisingly, that we are still prior dominated for multiple parameters, where our prior is simply the range of values considered in the LH simulations and thus the widest range of values our model would be valid over. A large part of why we are prior dominated is due to the sample size of available data, which is primarily from older surveys. This prior-domination can also be a sign of a non-optimal model; the 4 parameters varied in \textsc{Camels} were not optimized to find processes that most impact $\cvir$. While most of them \textit{do} have a strong impact on $\cvir$, we showed in Figure \ref{fig:TNG_SIMBA_slopes} that one of the AGN parameters had weak impact on $\cvir$, and this lends itself to the resulting posteriors being prior-dominated.

Even though many of our constraints are prior-dominated, we are still able to discuss qualitative behaviors for most parameters. Similar to our discussion on the simulation ``translation'' analysis, we re-emphasize here that our conservative claims are on whether the constraints prefer the feedback parameters to be larger/smaller than their fiducial value. A precise estimate of the value they must take is both not achievable given our data quality, and will also require significant, additional robustness checks that are beyond the scope of this work. Many of the approximations used for the observational data in this work (see Section \ref{sec:ObsData} for details) are inaccurate and thus will need to be changed to enable precise estimates.

Figure \ref{fig:ObsConstraints} shows our observational constraints. We find $\{A, B\}_{\rm SN1}$ prefers lower values than the fiducial, though we do not actually see a constraint here as the results are cutoff by the lower prior boundary. In TNG this reduces the energy scale of the SN feedback, and in SIMBA it reduces the mass-loading factor, which at fixed wind velocity reduces the SN energy scale as well. We do find interesting constraints on $\{A, B\}_{\rm SN2}$, which has a stronger impact on $\cvir$ and at a slightly higher halo mass than $\{A, B\}_{\rm SN1}$; TNG (SIMBA) appears to prefer values of $A_{\rm SN2}$ ($B_{\rm SN2}$) that are the same (higher) than the fiducial. This is consistent with our previous statement in Section \ref{sec:SimTranslation} that the stellar feedback in TNG, which causes the downturn in $\cvir-\Mvir$ at lower masses, is stronger than that in SIMBA. For the first AGN parameter, we see that TNG (SIMBA) appears to prefer values of $A_{\rm AGN1}$ ($B_{\rm AGN1}$) that are the lower (higher) than the fiducial value, while for the second one (which controls jet speed) both simulations prefer higher values than the fiducial. We also note that the $\chi^2$ of the results show an equal preference for TNG and SIMBA, which is expected given that the largest differences between the two simulation predictions occur at the lowest halo masses, where we do not have any constraining data,

Finally, one may worry to what extent other effects on the $\cvir-\Mvir$ relation (either physical or observational systematics-related) can appear as a feedback-related effect. In general, this is a very critical point to understand, especially if one were pursue precise, robust constraints on the parameters. Our model here can be thought of as a ``fiducial'' $\cvir-\Mvir$ relation from a hydrodynamic simulation, with additional ``basis'' functions --- in the form of the mass-dependent slopes --- to capture the impact of different parameters. Given these bases have a specific mass-dependence that is taken from simulations, a non-astrophysical effect will need to have a similar mass-dependence in order to mix their signal with an astrophysical effect.

Now, the caveat here is that feedback processes can still have a relatively mass-localized impact --- for example, \citet{Weinberger2018SMBHsIllustrisTNG} and \citetalias{Anbajagane2022BaryImprint} show the strongest impact for a given process (SN feedback, thermal AGN feedback etc.) is localized within $\sim 1\rm \dex$ in halo mass --- 
and so it is possible that any systematic that affects just halos at the lowest or highest masses (where the measurements come from different samples, and thus have different systematics) could more easily contaminate the astrophysical constraints. Both better data quality and a dedicated analysis of any effects would be required to disentagle all these factors. The purpose of this section is to demonstrate and place some observational constraints and then qualitatively interpret their meaning (i.e. are the parameter values higher/lower than the chosen fiducial values in TNG and SIMBA), so a detailed exploration is beyond the scope of our work.

A simple forecast -- where we take the mean $\cvir-\Mvir$ prediction from our TNG model with fiducial parameter values as the simulated datavector, add a 5\% measurement uncertainty, and run the MCMC --- shows that we can get constraints of around $\sigma(A_X) \in \{0.25, 1.0, 0.1, 0.3\}$ for the four astrophysical parameters. The median uncertainty of the existing measurements in Figure \ref{fig:ObsConstraints} is $\sim 20\%$.

\section{Conclusion}\label{sec:Conclusion}

The halo concentration--mass relation is a key quantity for understanding the formation and evolution of halos, and is actively being used in modelling many astrophysical and cosmological observables. Existing works have provided semi-analytic models for this relation as a function of cosmology using high-resolution DMO simulations. However, the quantitative impact of astrophysics on this relation is relatively unexplored, though many qualitative analyses already exist.

In this work, we use the \textsc{Camels} simulation suite to study the mass- and redshift-dependent response of halo concentration from galaxy formation processes, in particular SN and AGN feedback. We do so by building a locally linear but globally non-linear model of $\cvir$ as a function of redshift, mass, and cosmology/astrophysics parameters for two different galaxy formation models. Our choice allows for easy interpretability of our results, and the identification of physical processes leading to specific features, while capturing most of the nonlinearity in the relations. By using an uber halo sample constructed from 1000 simulations, we achieve higher statistical power and thereby analyze higher mass-scales that are normally unachievable in the small $L = 25 \hinv\mpc$ boxes of the individual \textsc{Camels} simulations.

Our main results are summarized as follows:
\begin{itemize}
    \item Figure \ref{fig:TNG_SIMBA_slopes} shows the dependence of $\cvir$ on various quantities. The SN feedback in TNG is most impactful at the $10^{11} - 10^{12} \msol/h$ mass range, given the shallower halo potentials at lower masses, and works to make the halo more diffuse. There continues to be some minor impact at larger mass scales, where gas cooling is known to be comparable/larger than stellar feedback. This is likely caused by changes to the mass loading factor --- controlled by both SN parameters --- which impacts the gas metal enrichment and thus metal cooling, which alters how efficiently the halo potential can contract. 
    \item AGN feedback in TNG becomes prominent at $\Mvir > 10^{13} \msol/h$, which is the mass at which the energy scale of kinetic feedback (radio jets) starts dominating the halo \citepalias{Anbajagane2022BaryImprint}.
    \item Figure \ref{fig:TNG_SIMBA_slopes} also shows that the response of $\cvir$ to SN and AGN feedback is quantitatively different in SIMBA compared to the response in TNG. This is expected as the equations being modified by the \textsc{Camels} astrophysics parameters are different. Notably, the SN mechanism is significantly weaker in SIMBA (reflected in how the slopes are closer to 0).
    \item The dependence of $\cvir$ on cosmology in the two simulations is generally consistent but differs in detail, especially at low masses. This highlights how the cosmology dependence, when conditioned on a given galaxy formation model, can have notable variations. The dependence on $\sigma_8$ has the same mass-dependence in all three of TNG, DMO, and semi-analytical predictions (Figure \ref{fig:TNG_DMO_Colossus}). The dependence on $\Omega_{\rm m}$ and $\Mvir$ is also similar but has notable non-monotonic deviation. 
    \item The redshift dependence of the \textit{qualitative} statements above is generally weak (Figure \ref{fig:slopes_z_dep}). Some interesting features are that the slope between $\cvir - A_{\rm SN1}$ shows a zero-crossing at low redshift, which then disappears at higher redshift. The slopes for $\cvir - A_{\rm SN2}$ also has a stationary point around $\Mvir = 5\times10^{12}\msol/h$, with strong redshift dependence below this mass-scale and weak dependence above it.
    \item Using our model to ``translate'' between simulations we can check how the \textsc{Camels} astrophysics parameters must be changed in order to get galaxy formation model \textbf{A} to match the fiducial result of model \textbf{B}. We find either that TNG must have lower SN feedback energy output, or that SIMBA must have lower mass loading and/or higher SN wind speeds. There is also a mild preference in TNG to have less bursty AGN kinetic jets.
    \item Finally, we do a simplistic test of our model on observational data for the $\cvir -\Mvir$ relation, and find that in most cases the astrophysical parameters prefer values away from the fiducial setting of $A_X = B_X = 1$ (Figure \ref{fig:ObsConstraints}). The parameters from the TNG- and SIMBA-model analysis imply a preference in both for having less energetic SN feedback and stronger AGN kinetic jet speeds.
\end{itemize}

Through performing a qualitative and quantitative study of $c_{\rm vir}$ in simulations, we extract the specific impact of galaxy formation on the matter distribution, identifying the mass- and redshift-dependent impact of specific processes. From a halo formation standpoint, this allows us to closely examine the exact impacts on the halo density distribution due to these processes. From an application standpoint, it allows us to understand the signature of specific processes on traditional, cosmological observables that can be forward-modelled using the halo model approach (eg. nonlinear power spectra).

Focusing on cosmological applications in particular, a highly relevant extension of this work would be to analyze a suite of simulations of a larger box size (while accommodating coarser mass resolution). Our main limitation in this work is the exact upper mass limit we can use in our scaling relations, and the cosmic variance component at these higher masses. While the current setup still allows for meaningful studies of fields like the total matter or lensing fields, which are most sensitive to halo masses well within our range, other studies of structure may require model predictions for more massive halos.

An ideal scenario of this type of work would also be to recover the DMO result by ``turning off'' the galaxy formation processes in a model for $\cvir-\Mvir$. However, this is possible only if we can turn off \textit{all} baryonic processes, which would require extending \textsc{Camels} to include on the order of $\mathcal{O}(100)$ parameters, rather than the 4 being currently considered. This is computationally infeasible at the moment. However, increasing the number of parameters by reasonable levels will continue to give us greater insight into the physics of galaxy formation and its impact on the larger structures.

The matter distribution within halos is an interesting problem. It has a rich phenomenology involving processes from gravitational collapse to gas cooling and feedback to metal enrichment to baryon backreactions and so forth. Understanding these processes is on its own interesting from a galaxy formation and galaxy evolution perspective, but it is then also a necessary component in understanding how the observed structure in our Universe behaves on nonlinear scales. Existing work has already extracted the qualitative features of baryon imprints. This work takes the first step in performing a more quantitative analysis of the same question, by leveraging a large number of simulations and answering how those baryon imprints vary quantitatively as specific galaxy formation processes are altered. As more observational constraints are obtained and more galaxy formation models are refined (through quantitative predictive models such as the ones explored here), we will be able to narrow the possible galaxy formation processes that take place in our Universe and have a more definitive picture of how baryons impact the matter distributions.

\section*{Acknowledgements}

We thank Andrew Hearin for many helpful comments on an earlier version of this draft. We also thank the \textsc{Camels} collaboration for making their data publicly available and easily accessible to the community. We thank the referee for their helpful comments.

DA is supported by the National Science Foundation Graduate Research Fellowship under Grant No. DGE 1746045. CC is supported by the Henry Luce Foundation and DOE grant DE-SC0021949.

All analysis in this work was enabled greatly by the following software: \textsc{Pandas} \citep{Mckinney2011pandas}, \textsc{NumPy} \citep{vanderWalt2011Numpy}, \textsc{SciPy} \citep{Virtanen2020Scipy}, and \textsc{Matplotlib} \citep{Hunter2007Matplotlib}. We have also used
the Astrophysics Data Service (\href{https://ui.adsabs.harvard.edu/}{ADS}) and \href{https://arxiv.org/}{\texttt{arXiv}} preprint repository extensively during this project and the writing of the paper.

\section*{Data Availability}

The tabulated \textsc{Kllr} parameters and the interpolator framework used in this work are made available at \url{https://github.com/mufanshao/CAMELSconcentration}. The public \textsc{Camels} data, including the \textsc{Rockstar} catalogs we use in this work, can be found at \url{https://camels.readthedocs.io}.

\bibliographystyle{mnras}
\bibliography{References}

\begin{thebibliography}{}
\makeatletter
\relax
\def\mn@urlcharsother{\let\do\@makeother \do\$\do\&\do\#\do\^\do\_\do\%\do\~}
\def\mn@doi{\begingroup\mn@urlcharsother \@ifnextchar [ {\mn@doi@}
  {\mn@doi@[]}}
\def\mn@doi@[#1]#2{\def\@tempa{#1}\ifx\@tempa\@empty \href
  {http://dx.doi.org/#2} {doi:#2}\else \href {http://dx.doi.org/#2} {#1}\fi
  \endgroup}
\def\mn@eprint#1#2{\mn@eprint@#1:#2::\@nil}
\def\mn@eprint@arXiv#1{\href {http://arxiv.org/abs/#1} {{\tt arXiv:#1}}}
\def\mn@eprint@dblp#1{\href {http://dblp.uni-trier.de/rec/bibtex/#1.xml}
  {dblp:#1}}
\def\mn@eprint@#1:#2:#3:#4\@nil{\def\@tempa {#1}\def\@tempb {#2}\def\@tempc
  {#3}\ifx \@tempc \@empty \let \@tempc \@tempb \let \@tempb \@tempa \fi \ifx
  \@tempb \@empty \def\@tempb {arXiv}\fi \@ifundefined
  {mn@eprint@\@tempb}{\@tempb:\@tempc}{\expandafter \expandafter \csname
  mn@eprint@\@tempb\endcsname \expandafter{\@tempc}}}

\bibitem[\protect\citeauthoryear{{Abadi}, {Navarro}, {Fardal}, {Babul}  \&
  {Steinmetz}}{{Abadi} et~al.}{2010}]{Abadi2010ShapesBaryons}
{Abadi} M.~G.,  {Navarro} J.~F.,  {Fardal} M.,  {Babul} A.,   {Steinmetz} M.,
  2010, \mn@doi [\mnras] {10.1111/j.1365-2966.2010.16912.x}, \href
  {https://ui.adsabs.harvard.edu/abs/2010MNRAS.407..435A} {407, 435}

\bibitem[\protect\citeauthoryear{{Abbott} et~al.,}{{Abbott}
  et~al.}{2022}]{DES2022Y3}
{Abbott} T.~M.~C.,  et~al., 2022, \mn@doi [\prd] {10.1103/PhysRevD.105.023520},
  \href {https://ui.adsabs.harvard.edu/abs/2022PhRvD.105b3520A} {105, 023520}

\bibitem[\protect\citeauthoryear{{Allen}, {Evrard}  \& {Mantz}}{{Allen}
  et~al.}{2011}]{Allen2011CosmoClusterReview}
{Allen} S.~W.,  {Evrard} A.~E.,   {Mantz} A.~B.,  2011, \mn@doi [\araa]
  {10.1146/annurev-astro-081710-102514}, \href
  {https://ui.adsabs.harvard.edu/abs/2011ARA&A..49..409A} {49, 409}

\bibitem[\protect\citeauthoryear{{Amon} et~al.,}{{Amon}
  et~al.}{2022}]{Amon2022Y3shear}
{Amon} A.,  et~al., 2022, \mn@doi [\prd] {10.1103/PhysRevD.105.023514}, \href
  {https://ui.adsabs.harvard.edu/abs/2022PhRvD.105b3514A} {105, 023514}

\bibitem[\protect\citeauthoryear{{Anbajagane}, {Evrard}, {Farahi}, {Barnes},
  {Dolag}, {McCarthy}, {Nelson}  \& {Pillepich}}{{Anbajagane}
  et~al.}{2020}]{Anbajagane2020StellarStatistics}
{Anbajagane} D.,  {Evrard} A.~E.,  {Farahi} A.,  {Barnes} D.~J.,  {Dolag} K.,
  {McCarthy} I.~G.,  {Nelson} D.,   {Pillepich} A.,  2020, \mn@doi [\mnras]
  {10.1093/mnras/staa1147}, \href
  {https://ui.adsabs.harvard.edu/abs/2020MNRAS.495..686A} {495, 686}

\bibitem[\protect\citeauthoryear{{Anbajagane}, {Evrard}  \&
  {Farahi}}{{Anbajagane} et~al.}{2022a}]{Anbajagane2022BaryImprint}
{Anbajagane} D.,  {Evrard} A.~E.,   {Farahi} A.,  2022a, \mn@doi [\mnras]
  {10.1093/mnras/stab3177}, \href
  {https://ui.adsabs.harvard.edu/abs/2022MNRAS.509.3441A} {509, 3441}

\bibitem[\protect\citeauthoryear{{Anbajagane} et~al.,}{{Anbajagane}
  et~al.}{2022b}]{Anbajagane2022GalVelBias}
{Anbajagane} D.,  et~al., 2022b, \mn@doi [\mnras] {10.1093/mnras/stab3587},
  \href {https://ui.adsabs.harvard.edu/abs/2022MNRAS.510.2980A} {510, 2980}

\bibitem[\protect\citeauthoryear{{Anbajagane} et~al.,}{{Anbajagane}
  et~al.}{2022c}]{Anbajagane2022Shocks}
{Anbajagane} D.,  et~al., 2022c, \mn@doi [\mnras] {10.1093/mnras/stac1376},
  \href {https://ui.adsabs.harvard.edu/abs/2022MNRAS.514.1645A} {514, 1645}

\bibitem[\protect\citeauthoryear{{Artale}, {Pedrosa}, {Tissera}, {Cataldi}  \&
  {Di Cintio}}{{Artale} et~al.}{2019}]{Artale2019BaryonImprints}
{Artale} M.~C.,  {Pedrosa} S.~E.,  {Tissera} P.~B.,  {Cataldi} P.,   {Di
  Cintio} A.,  2019, \mn@doi [\aap] {10.1051/0004-6361/201834096}, \href
  {https://ui.adsabs.harvard.edu/abs/2019A&A...622A.197A} {622, A197}

\bibitem[\protect\citeauthoryear{{Asgari} et~al.,}{{Asgari}
  et~al.}{2021}]{Asgari2021KidsWL}
{Asgari} M.,  et~al., 2021, \mn@doi [\aap] {10.1051/0004-6361/202039070}, \href
  {https://ui.adsabs.harvard.edu/abs/2021A&A...645A.104A} {645, A104}

\bibitem[\protect\citeauthoryear{{Ayromlou}, {Nelson}  \&
  {Pillepich}}{{Ayromlou} et~al.}{2022}]{Ayromlou2022ClosureRadius}
{Ayromlou} M.,  {Nelson} D.,   {Pillepich} A.,  2022, arXiv e-prints, \href
  {https://ui.adsabs.harvard.edu/abs/2022arXiv221107659A} {p. arXiv:2211.07659}

\bibitem[\protect\citeauthoryear{{Behroozi}, {Wechsler}  \& {Wu}}{{Behroozi}
  et~al.}{2013a}]{Behroozi2013Rockstar}
{Behroozi} P.~S.,  {Wechsler} R.~H.,   {Wu} H.-Y.,  2013a, \mn@doi [\apj]
  {10.1088/0004-637X/762/2/109}, \href
  {https://ui.adsabs.harvard.edu/abs/2013ApJ...762..109B} {762, 109}

\bibitem[\protect\citeauthoryear{{Behroozi}, {Wechsler}  \&
  {Conroy}}{{Behroozi} et~al.}{2013b}]{Behroozi2013StarFormationHistory}
{Behroozi} P.~S.,  {Wechsler} R.~H.,   {Conroy} C.,  2013b, \mn@doi [\apj]
  {10.1088/0004-637X/770/1/57}, \href
  {https://ui.adsabs.harvard.edu/abs/2013ApJ...770...57B} {770, 57}

\bibitem[\protect\citeauthoryear{{Beltz-Mohrmann} \&
  {Berlind}}{{Beltz-Mohrmann} \&
  {Berlind}}{2021}]{Beltz-Mohrmann2021BaryonImpactTNG}
{Beltz-Mohrmann} G.~D.,  {Berlind} A.~A.,  2021, arXiv e-prints, \href
  {https://ui.adsabs.harvard.edu/abs/2021arXiv210305076B} {p. arXiv:2103.05076}

\bibitem[\protect\citeauthoryear{{Bhattacharya}, {Habib}, {Heitmann}  \&
  {Vikhlinin}}{{Bhattacharya} et~al.}{2013}]{Bhattacharya2013Concentrations}
{Bhattacharya} S.,  {Habib} S.,  {Heitmann} K.,   {Vikhlinin} A.,  2013,
  \mn@doi [\apj] {10.1088/0004-637X/766/1/32}, \href
  {https://ui.adsabs.harvard.edu/abs/2013ApJ...766...32B} {766, 32}

\bibitem[\protect\citeauthoryear{{Blumenthal}, {Faber}, {Flores}  \&
  {Primack}}{{Blumenthal} et~al.}{1986}]{Blumenthal1986AdiabaticContraction}
{Blumenthal} G.~R.,  {Faber} S.~M.,  {Flores} R.,   {Primack} J.~R.,  1986,
  \mn@doi [\apj] {10.1086/163867}, \href
  {https://ui.adsabs.harvard.edu/abs/1986ApJ...301...27B} {301, 27}

\bibitem[\protect\citeauthoryear{{Brimioulle}, {Seitz}, {Lerchster}, {Bender}
  \& {Snigula}}{{Brimioulle} et~al.}{2013}]{Brimioulle2013CFHTCvir}
{Brimioulle} F.,  {Seitz} S.,  {Lerchster} M.,  {Bender} R.,   {Snigula} J.,
  2013, \mn@doi [\mnras] {10.1093/mnras/stt525}, \href
  {https://ui.adsabs.harvard.edu/abs/2013MNRAS.432.1046B} {432, 1046}

\bibitem[\protect\citeauthoryear{{Bryan} \& {Norman}}{{Bryan} \&
  {Norman}}{1998}]{Bryan1998vir}
{Bryan} G.~L.,  {Norman} M.~L.,  1998, \mn@doi [\apj] {10.1086/305262}, \href
  {https://ui.adsabs.harvard.edu/abs/1998ApJ...495...80B} {495, 80}

\bibitem[\protect\citeauthoryear{{Bryan}, {Kay}, {Duffy}, {Schaye}, {Dalla
  Vecchia}  \& {Booth}}{{Bryan} et~al.}{2013}]{Bryan2013BaryonImpactOnShapes}
{Bryan} S.~E.,  {Kay} S.~T.,  {Duffy} A.~R.,  {Schaye} J.,  {Dalla Vecchia} C.,
    {Booth} C.~M.,  2013, \mn@doi [\mnras] {10.1093/mnras/sts587}, \href
  {https://ui.adsabs.harvard.edu/abs/2013MNRAS.429.3316B} {429, 3316}

\bibitem[\protect\citeauthoryear{{Child}, {Habib}, {Heitmann}, {Frontiere},
  {Finkel}, {Pope}  \& {Morozov}}{{Child}
  et~al.}{2018}]{Child2018ConcentratioMassRelation}
{Child} H.~L.,  {Habib} S.,  {Heitmann} K.,  {Frontiere} N.,  {Finkel} H.,
  {Pope} A.,   {Morozov} V.,  2018, \mn@doi [\apj] {10.3847/1538-4357/aabf95},
  \href {https://ui.adsabs.harvard.edu/abs/2018ApJ...859...55C} {859, 55}

\bibitem[\protect\citeauthoryear{{Chisari} et~al.,}{{Chisari}
  et~al.}{2018}]{Chisari2018BaryonsPk}
{Chisari} N.~E.,  et~al., 2018, \mn@doi [\mnras] {10.1093/mnras/sty2093}, \href
  {https://ui.adsabs.harvard.edu/abs/2018MNRAS.480.3962C} {480, 3962}

\bibitem[\protect\citeauthoryear{{Chua}, {Pillepich}, {Vogelsberger}  \&
  {Hernquist}}{{Chua} et~al.}{2019}]{Chua2019ShapeIllustrisBaryons}
{Chua} K. T.~E.,  {Pillepich} A.,  {Vogelsberger} M.,   {Hernquist} L.,  2019,
  \mn@doi [\mnras] {10.1093/mnras/sty3531}, \href
  {https://ui.adsabs.harvard.edu/abs/2019MNRAS.484..476C} {484, 476}

\bibitem[\protect\citeauthoryear{{Chua}, {Vogelsberger}, {Pillepich}  \&
  {Hernquist}}{{Chua} et~al.}{2021}]{Chua2021TNGShapes}
{Chua} K. T.~E.,  {Vogelsberger} M.,  {Pillepich} A.,   {Hernquist} L.,  2021,
  arXiv e-prints, \href {https://ui.adsabs.harvard.edu/abs/2021arXiv210900012C}
  {p. arXiv:2109.00012}

\bibitem[\protect\citeauthoryear{{Cooray} \& {Sheth}}{{Cooray} \&
  {Sheth}}{2002}]{Cooray2002HaloModel}
{Cooray} A.,  {Sheth} R.,  2002, \mn@doi [\physrep]
  {10.1016/S0370-1573(02)00276-4}, \href
  {https://ui.adsabs.harvard.edu/abs/2002PhR...372....1C} {372, 1}

\bibitem[\protect\citeauthoryear{{Cui} et~al.,}{{Cui}
  et~al.}{2016}]{Cui2016NiftyBaryonsHaloProperties}
{Cui} W.,  et~al., 2016, \mn@doi [\mnras] {10.1093/mnras/stw603}, \href
  {https://ui.adsabs.harvard.edu/abs/2016MNRAS.458.4052C} {458, 4052}

\bibitem[\protect\citeauthoryear{{Cui} et~al.,}{{Cui}
  et~al.}{2018}]{Cui2018The300}
{Cui} W.,  et~al., 2018, \mn@doi [\mnras] {10.1093/mnras/sty2111}, \href
  {https://ui.adsabs.harvard.edu/abs/2018MNRAS.480.2898C} {480, 2898}

\bibitem[\protect\citeauthoryear{{Cui} et~al.,}{{Cui}
  et~al.}{2022}]{Cui2022SIMBA}
{Cui} W.,  et~al., 2022, \mn@doi [\mnras] {10.1093/mnras/stac1402}, \href
  {https://ui.adsabs.harvard.edu/abs/2022MNRAS.514..977C} {514, 977}

\bibitem[\protect\citeauthoryear{{Dav{\'e}}, {Angl{\'e}s-Alc{\'a}zar},
  {Narayanan}, {Li}, {Rafieferantsoa}  \& {Appleby}}{{Dav{\'e}}
  et~al.}{2019}]{Dave2019SIMBA}
{Dav{\'e}} R.,  {Angl{\'e}s-Alc{\'a}zar} D.,  {Narayanan} D.,  {Li} Q.,
  {Rafieferantsoa} M.~H.,   {Appleby} S.,  2019, \mn@doi [\mnras]
  {10.1093/mnras/stz937}, \href
  {https://ui.adsabs.harvard.edu/abs/2019MNRAS.486.2827D} {486, 2827}

\bibitem[\protect\citeauthoryear{{Diemer}}{{Diemer}}{2018}]{Diemer2018COLOSSUS}
{Diemer} B.,  2018, \mn@doi [\apjs] {10.3847/1538-4365/aaee8c}, \href
  {https://ui.adsabs.harvard.edu/abs/2018ApJS..239...35D} {239, 35}

\bibitem[\protect\citeauthoryear{{Diemer} \& {Joyce}}{{Diemer} \&
  {Joyce}}{2019}]{Diemer2019concentrations}
{Diemer} B.,  {Joyce} M.,  2019, \mn@doi [\apj] {10.3847/1538-4357/aafad6},
  \href {https://ui.adsabs.harvard.edu/abs/2019ApJ...871..168D} {871, 168}

\bibitem[\protect\citeauthoryear{{Diemer} \& {Kravtsov}}{{Diemer} \&
  {Kravtsov}}{2015}]{Diemer2015Concentration}
{Diemer} B.,  {Kravtsov} A.~V.,  2015, \mn@doi [\apj]
  {10.1088/0004-637X/799/1/108}, \href
  {https://ui.adsabs.harvard.edu/abs/2015ApJ...799..108D} {799, 108}

\bibitem[\protect\citeauthoryear{{Duffy}, {Schaye}, {Kay}, {Dalla Vecchia},
  {Battye}  \& {Booth}}{{Duffy} et~al.}{2010}]{Duffy2010BaryonDmProfileDensity}
{Duffy} A.~R.,  {Schaye} J.,  {Kay} S.~T.,  {Dalla Vecchia} C.,  {Battye}
  R.~A.,   {Booth} C.~M.,  2010, \mn@doi [\mnras]
  {10.1111/j.1365-2966.2010.16613.x}, \href
  {https://ui.adsabs.harvard.edu/abs/2010MNRAS.405.2161D} {405, 2161}

\bibitem[\protect\citeauthoryear{{Farahi}, {Evrard}, {McCarthy}, {Barnes}  \&
  {Kay}}{{Farahi} et~al.}{2018}]{Farahi2018BAHAMAS}
{Farahi} A.,  {Evrard} A.~E.,  {McCarthy} I.,  {Barnes} D.~J.,   {Kay} S.~T.,
  2018, \mn@doi [\mnras] {10.1093/mnras/sty1179}, \href
  {https://ui.adsabs.harvard.edu/abs/2018MNRAS.478.2618F} {478, 2618}

\bibitem[\protect\citeauthoryear{{Farahi}, {Nagai}  \& {Chen}}{{Farahi}
  et~al.}{2021}]{Farahi2021PoPE}
{Farahi} A.,  {Nagai} D.,   {Chen} Y.,  2021, \mn@doi [\aj]
  {10.3847/1538-3881/abc630}, \href
  {https://ui.adsabs.harvard.edu/abs/2021AJ....161...30F} {161, 30}

\bibitem[\protect\citeauthoryear{{Farahi}, {Nagai}  \& {Anbajagane}}{{Farahi}
  et~al.}{2022a}]{Farahi2022ProfileCorr}
{Farahi} A.,  {Nagai} D.,   {Anbajagane} D.,  2022a, arXiv e-prints, \href
  {https://ui.adsabs.harvard.edu/abs/2022arXiv220413578F} {p. arXiv:2204.13578}

\bibitem[\protect\citeauthoryear{{Farahi}, {Anbajagane}  \& {Evrard}}{{Farahi}
  et~al.}{2022b}]{Farahi2022KLLR}
{Farahi} A.,  {Anbajagane} D.,   {Evrard} A.~E.,  2022b, \mn@doi [\apj]
  {10.3847/1538-4357/ac6ac7}, \href
  {https://ui.adsabs.harvard.edu/abs/2022ApJ...931..166F} {931, 166}

\bibitem[\protect\citeauthoryear{{Foreman-Mackey}, {Hogg}, {Lang}  \&
  {Goodman}}{{Foreman-Mackey} et~al.}{2013}]{Foreman-Mackey2013emcee}
{Foreman-Mackey} D.,  {Hogg} D.~W.,  {Lang} D.,   {Goodman} J.,  2013, \mn@doi
  [\pasp] {10.1086/670067}, \href
  {https://ui.adsabs.harvard.edu/abs/2013PASP..125..306F} {125, 306}

\bibitem[\protect\citeauthoryear{{Forouhar Moreno}, {Ben{\'\i}tez-Llambay},
  {Cole}  \& {Frenk}}{{Forouhar Moreno}
  et~al.}{2021}]{ForouharMoreno2021BaryonsConcentrationEagle}
{Forouhar Moreno} V.~J.,  {Ben{\'\i}tez-Llambay} A.,  {Cole} S.,   {Frenk} C.,
  2021, arXiv e-prints, \href
  {https://ui.adsabs.harvard.edu/abs/2021arXiv210714245F} {p. arXiv:2107.14245}

\bibitem[\protect\citeauthoryear{{Gatti} et~al.,}{{Gatti}
  et~al.}{2021}]{Gatti2021DESxACT}
{Gatti} M.,  et~al., 2021, arXiv e-prints, \href
  {https://ui.adsabs.harvard.edu/abs/2021arXiv210801600G} {p. arXiv:2108.01600}

\bibitem[\protect\citeauthoryear{{Gnedin}, {Kravtsov}, {Klypin}  \&
  {Nagai}}{{Gnedin} et~al.}{2004}]{Gnedin2004AdiabaticContraction}
{Gnedin} O.~Y.,  {Kravtsov} A.~V.,  {Klypin} A.~A.,   {Nagai} D.,  2004,
  \mn@doi [\apj] {10.1086/424914}, \href
  {https://ui.adsabs.harvard.edu/abs/2004ApJ...616...16G} {616, 16}

\bibitem[\protect\citeauthoryear{{Heitmann}, {Higdon}, {White}, {Habib},
  {Williams}, {Lawrence}  \& {Wagner}}{{Heitmann}
  et~al.}{2009}]{Heitmann2009Emulator}
{Heitmann} K.,  {Higdon} D.,  {White} M.,  {Habib} S.,  {Williams} B.~J.,
  {Lawrence} E.,   {Wagner} C.,  2009, \mn@doi [\apj]
  {10.1088/0004-637X/705/1/156}, \href
  {https://ui.adsabs.harvard.edu/abs/2009ApJ...705..156H} {705, 156}

\bibitem[\protect\citeauthoryear{{Hirschmann}, {Dolag}, {Saro}, {Bachmann},
  {Borgani}  \& {Burkert}}{{Hirschmann} et~al.}{2014}]{Hirschmann2014MGTM}
{Hirschmann} M.,  {Dolag} K.,  {Saro} A.,  {Bachmann} L.,  {Borgani} S.,
  {Burkert} A.,  2014, \mn@doi [\mnras] {10.1093/mnras/stu1023}, \href
  {https://ui.adsabs.harvard.edu/abs/2014MNRAS.442.2304H} {442, 2304}

\bibitem[\protect\citeauthoryear{{Hu} \& {Kravtsov}}{{Hu} \&
  {Kravtsov}}{2003}]{Hu2003SampleVar}
{Hu} W.,  {Kravtsov} A.~V.,  2003, \mn@doi [\apj] {10.1086/345846}, \href
  {https://ui.adsabs.harvard.edu/abs/2003ApJ...584..702H} {584, 702}

\bibitem[\protect\citeauthoryear{{Hunter}}{{Hunter}}{2007}]{Hunter2007Matplotlib}
{Hunter} J.~D.,  2007, \mn@doi [Computing in Science and Engineering]
  {10.1109/MCSE.2007.55}, \href
  {https://ui.adsabs.harvard.edu/abs/2007CSE.....9...90H} {9, 90}

\bibitem[\protect\citeauthoryear{{Ishiyama} et~al.,}{{Ishiyama}
  et~al.}{2021}]{Ishiyama2020UchuuConcentration}
{Ishiyama} T.,  et~al., 2021, \mn@doi [\mnras] {10.1093/mnras/stab1755}, \href
  {https://ui.adsabs.harvard.edu/abs/2021MNRAS.506.4210I} {506, 4210}

\bibitem[\protect\citeauthoryear{{Johnston} et~al.,}{{Johnston}
  et~al.}{2007}]{Johnston2007Concentration}
{Johnston} D.~E.,  et~al., 2007, arXiv e-prints, \href
  {https://ui.adsabs.harvard.edu/abs/2007arXiv0709.1159J} {p. arXiv:0709.1159}

\bibitem[\protect\citeauthoryear{{Kravtsov} \& {Borgani}}{{Kravtsov} \&
  {Borgani}}{2012}]{Kravtsov2012ClusterFormation}
{Kravtsov} A.~V.,  {Borgani} S.,  2012, \mn@doi [\araa]
  {10.1146/annurev-astro-081811-125502}, \href
  {https://ui.adsabs.harvard.edu/abs/2012ARA&A..50..353K} {50, 353}

\bibitem[\protect\citeauthoryear{{Lee} et~al.,}{{Lee}
  et~al.}{2022}]{Lee2022rSZ}
{Lee} E.,  et~al., 2022, \mn@doi [\mnras] {10.1093/mnras/stac2781}, \href
  {https://ui.adsabs.harvard.edu/abs/2022MNRAS.517.5303L} {517, 5303}

\bibitem[\protect\citeauthoryear{{Machado Poletti Valle}, {Avestruz}, {Barnes},
  {Farahi}, {Lau}  \& {Nagai}}{{Machado Poletti Valle}
  et~al.}{2020}]{Machado2020GasShapesSHAP}
{Machado Poletti Valle} L.~F.,  {Avestruz} C.,  {Barnes} D.~J.,  {Farahi} A.,
  {Lau} E.~T.,   {Nagai} D.,  2020, arXiv e-prints, \href
  {https://ui.adsabs.harvard.edu/abs/2020arXiv201112987M} {p. arXiv:2011.12987}

\bibitem[\protect\citeauthoryear{{Mandelbaum}, {Seljak}, {Cool}, {Blanton},
  {Hirata}  \& {Brinkmann}}{{Mandelbaum} et~al.}{2006}]{Mandelbaum2006SDSSCvir}
{Mandelbaum} R.,  {Seljak} U.,  {Cool} R.~J.,  {Blanton} M.,  {Hirata} C.~M.,
  {Brinkmann} J.,  2006, \mn@doi [\mnras] {10.1111/j.1365-2966.2006.10906.x},
  \href {https://ui.adsabs.harvard.edu/abs/2006MNRAS.372..758M} {372, 758}

\bibitem[\protect\citeauthoryear{{Mandelbaum}, {Seljak}  \&
  {Hirata}}{{Mandelbaum} et~al.}{2008}]{Mandelbaum2008ConcentrationMeasurement}
{Mandelbaum} R.,  {Seljak} U.,   {Hirata} C.~M.,  2008, \mn@doi [\jcap]
  {10.1088/1475-7516/2008/08/006}, \href
  {https://ui.adsabs.harvard.edu/abs/2008JCAP...08..006M} {2008, 006}

\bibitem[\protect\citeauthoryear{{Marinacci} et~al.,}{{Marinacci}
  et~al.}{2018}]{Marinacci2018FirstFields}
{Marinacci} F.,  et~al., 2018, \mn@doi [\mnras] {10.1093/mnras/sty2206}, \href
  {https://ui.adsabs.harvard.edu/abs/2018MNRAS.480.5113M} {480, 5113}

\bibitem[\protect\citeauthoryear{{McCarthy}, {Schaye}, {Bird}  \& {Le
  Brun}}{{McCarthy} et~al.}{2017}]{McCarthy2017BAHAMAS}
{McCarthy} I.~G.,  {Schaye} J.,  {Bird} S.,   {Le Brun} A. M.~C.,  2017,
  \mn@doi [\mnras] {10.1093/mnras/stw2792}, \href
  {https://ui.adsabs.harvard.edu/abs/2017MNRAS.465.2936M} {465, 2936}

\bibitem[\protect\citeauthoryear{McKinney}{McKinney}{2011}]{Mckinney2011pandas}
McKinney W.,  2011, Python for High Performance and Scientific Computing, 14

\bibitem[\protect\citeauthoryear{{Muratov}, {Kere{\v{s}}},
  {Faucher-Gigu{\`e}re}, {Hopkins}, {Quataert}  \& {Murray}}{{Muratov}
  et~al.}{2015}]{Muratov2015GasOutflowsFIRE}
{Muratov} A.~L.,  {Kere{\v{s}}} D.,  {Faucher-Gigu{\`e}re} C.-A.,  {Hopkins}
  P.~F.,  {Quataert} E.,   {Murray} N.,  2015, \mn@doi [\mnras]
  {10.1093/mnras/stv2126}, \href
  {https://ui.adsabs.harvard.edu/abs/2015MNRAS.454.2691M} {454, 2691}

\bibitem[\protect\citeauthoryear{{Naiman} et~al.,}{{Naiman}
  et~al.}{2018}]{Naiman2018FirstEuropium}
{Naiman} J.~P.,  et~al., 2018, \mn@doi [\mnras] {10.1093/mnras/sty618}, \href
  {https://ui.adsabs.harvard.edu/abs/2018MNRAS.477.1206N} {477, 1206}

\bibitem[\protect\citeauthoryear{{Navarro}, {Frenk}  \& {White}}{{Navarro}
  et~al.}{1997}]{Navarro1997NFWProfile}
{Navarro} J.~F.,  {Frenk} C.~S.,   {White} S. D.~M.,  1997, \mn@doi [\apj]
  {10.1086/304888}, \href
  {https://ui.adsabs.harvard.edu/abs/1997ApJ...490..493N} {490, 493}

\bibitem[\protect\citeauthoryear{{Nelson} et~al.,}{{Nelson}
  et~al.}{2018}]{Nelson2018FirstBimodality}
{Nelson} D.,  et~al., 2018, \mn@doi [\mnras] {10.1093/mnras/stx3040}, \href
  {https://ui.adsabs.harvard.edu/abs/2018MNRAS.475..624N} {475, 624}

\bibitem[\protect\citeauthoryear{{Nelson} et~al.,}{{Nelson}
  et~al.}{2019}]{Nelson2019TNG50}
{Nelson} D.,  et~al., 2019, \mn@doi [\mnras] {10.1093/mnras/stz2306}, \href
  {https://ui.adsabs.harvard.edu/abs/2019MNRAS.490.3234N} {490, 3234}

\bibitem[\protect\citeauthoryear{{Ntampaka} et~al.,}{{Ntampaka}
  et~al.}{2019}]{Ntampaka2019XrayClustersML}
{Ntampaka} M.,  et~al., 2019, \mn@doi [\apj] {10.3847/1538-4357/ab14eb}, \href
  {https://ui.adsabs.harvard.edu/abs/2019ApJ...876...82N} {876, 82}

\bibitem[\protect\citeauthoryear{{Pandey} et~al.,}{{Pandey}
  et~al.}{2019}]{Pandey2019GalaxytSZ}
{Pandey} S.,  et~al., 2019, \mn@doi [\prd] {10.1103/PhysRevD.100.063519}, \href
  {https://ui.adsabs.harvard.edu/abs/2019PhRvD.100f3519P} {100, 063519}

\bibitem[\protect\citeauthoryear{{Pandey} et~al.,}{{Pandey}
  et~al.}{2021}]{Pandey2021DESxACT}
{Pandey} S.,  et~al., 2021, arXiv e-prints, \href
  {https://ui.adsabs.harvard.edu/abs/2021arXiv210801601P} {p. arXiv:2108.01601}

\bibitem[\protect\citeauthoryear{{Pedrosa}, {Tissera}  \&
  {Scannapieco}}{{Pedrosa} et~al.}{2010}]{Pedrosa2010BaryonImprints}
{Pedrosa} S.,  {Tissera} P.~B.,   {Scannapieco} C.,  2010, \mn@doi [\mnras]
  {10.1111/j.1365-2966.2009.15951.x}, \href
  {https://ui.adsabs.harvard.edu/abs/2010MNRAS.402..776P} {402, 776}

\bibitem[\protect\citeauthoryear{{Pillepich} et~al.,}{{Pillepich}
  et~al.}{2018a}]{Pillepich2018Methods}
{Pillepich} A.,  et~al., 2018a, \mn@doi [\mnras] {10.1093/mnras/stx2656}, \href
  {https://ui.adsabs.harvard.edu/abs/2018MNRAS.473.4077P} {473, 4077}

\bibitem[\protect\citeauthoryear{{Pillepich} et~al.,}{{Pillepich}
  et~al.}{2018b}]{Pillepich2018FirstGalaxies}
{Pillepich} A.,  et~al., 2018b, \mn@doi [\mnras] {10.1093/mnras/stx3112}, \href
  {https://ui.adsabs.harvard.edu/abs/2018MNRAS.475..648P} {475, 648}

\bibitem[\protect\citeauthoryear{{Pillepich} et~al.,}{{Pillepich}
  et~al.}{2019}]{Pillepich2019TNG50}
{Pillepich} A.,  et~al., 2019, \mn@doi [\mnras] {10.1093/mnras/stz2338}, \href
  {https://ui.adsabs.harvard.edu/abs/2019MNRAS.490.3196P} {490, 3196}

\bibitem[\protect\citeauthoryear{{Planck Collaboration} et~al.,}{{Planck
  Collaboration} et~al.}{2016}]{Planck2015CosmoParams}
{Planck Collaboration} et~al., 2016, \mn@doi [\aap]
  {10.1051/0004-6361/201525830}, \href
  {https://ui.adsabs.harvard.edu/abs/2016A&A...594A..13P} {594, A13}

\bibitem[\protect\citeauthoryear{{Ragagnin}, {Dolag}, {Moscardini}, {Biviano}
  \& {D'Onofrio}}{{Ragagnin} et~al.}{2019}]{Ragagnin2019HaloConcentration}
{Ragagnin} A.,  {Dolag} K.,  {Moscardini} L.,  {Biviano} A.,   {D'Onofrio} M.,
  2019, \mn@doi [\mnras] {10.1093/mnras/stz1103}, \href
  {https://ui.adsabs.harvard.edu/abs/2019MNRAS.486.4001R} {486, 4001}

\bibitem[\protect\citeauthoryear{{Rudd}, {Zentner}  \& {Kravtsov}}{{Rudd}
  et~al.}{2008}]{Rudd2008BaryonsMPk}
{Rudd} D.~H.,  {Zentner} A.~R.,   {Kravtsov} A.~V.,  2008, \mn@doi [\apj]
  {10.1086/523836}, \href
  {https://ui.adsabs.harvard.edu/abs/2008ApJ...672...19R} {672, 19}

\bibitem[\protect\citeauthoryear{{S{\'a}nchez} et~al.,}{{S{\'a}nchez}
  et~al.}{2022}]{Sanchez2022Sheary}
{S{\'a}nchez} J.,  et~al., 2022, arXiv e-prints, \href
  {https://ui.adsabs.harvard.edu/abs/2022arXiv221008633S} {p. arXiv:2210.08633}

\bibitem[\protect\citeauthoryear{{Schneider}, {Teyssier}, {Stadel}, {Chisari},
  {Le Brun}, {Amara}  \& {Refregier}}{{Schneider}
  et~al.}{2019}]{Schneider2019BaryonsPk}
{Schneider} A.,  {Teyssier} R.,  {Stadel} J.,  {Chisari} N.~E.,  {Le Brun} A.
  M.~C.,  {Amara} A.,   {Refregier} A.,  2019, \mn@doi [\jcap]
  {10.1088/1475-7516/2019/03/020}, \href
  {https://ui.adsabs.harvard.edu/abs/2019JCAP...03..020S} {2019, 020}

\bibitem[\protect\citeauthoryear{{Schneider}, {Giri}, {Amodeo}  \&
  {Refregier}}{{Schneider} et~al.}{2022}]{Schneider2022KidsAstro}
{Schneider} A.,  {Giri} S.~K.,  {Amodeo} S.,   {Refregier} A.,  2022, \mn@doi
  [\mnras] {10.1093/mnras/stac1493}, \href
  {https://ui.adsabs.harvard.edu/abs/2022MNRAS.514.3802S} {514, 3802}

\bibitem[\protect\citeauthoryear{{Secco} et~al.,}{{Secco}
  et~al.}{2022}]{Secco2022Y3Shear}
{Secco} L.~F.,  et~al., 2022, \mn@doi [\prd] {10.1103/PhysRevD.105.023515},
  \href {https://ui.adsabs.harvard.edu/abs/2022PhRvD.105b3515S} {105, 023515}

\bibitem[\protect\citeauthoryear{{Semboloni}, {Hoekstra}, {Schaye}, {van
  Daalen}  \& {McCarthy}}{{Semboloni} et~al.}{2011}]{Sembolini2011BaryonsWL}
{Semboloni} E.,  {Hoekstra} H.,  {Schaye} J.,  {van Daalen} M.~P.,   {McCarthy}
  I.~G.,  2011, \mn@doi [\mnras] {10.1111/j.1365-2966.2011.19385.x}, \href
  {https://ui.adsabs.harvard.edu/abs/2011MNRAS.417.2020S} {417, 2020}

\bibitem[\protect\citeauthoryear{{Seppi} et~al.,}{{Seppi}
  et~al.}{2021}]{Seppi2021HMF}
{Seppi} R.,  et~al., 2021, \mn@doi [\aap] {10.1051/0004-6361/202039123}, \href
  {https://ui.adsabs.harvard.edu/abs/2021A&A...652A.155S} {652, A155}

\bibitem[\protect\citeauthoryear{{Shen}, {Xiao}, {Hopkins}  \& {Zurek}}{{Shen}
  et~al.}{2022}]{Shen2022AxionMini}
{Shen} X.,  {Xiao} H.,  {Hopkins} P.~F.,   {Zurek} K.~M.,  2022, arXiv
  e-prints, \href {https://ui.adsabs.harvard.edu/abs/2022arXiv220711276S} {p.
  arXiv:2207.11276}

\bibitem[\protect\citeauthoryear{{Shin} et~al.,}{{Shin}
  et~al.}{2021}]{Shin2021MassGalaxyProfilesDES}
{Shin} T.,  et~al., 2021, arXiv e-prints, \href
  {https://ui.adsabs.harvard.edu/abs/2021arXiv210505914S} {p. arXiv:2105.05914}

\bibitem[\protect\citeauthoryear{{Springel} et~al.,}{{Springel}
  et~al.}{2018}]{Springel2018FirstClustering}
{Springel} V.,  et~al., 2018, \mn@doi [\mnras] {10.1093/mnras/stx3304}, \href
  {https://ui.adsabs.harvard.edu/abs/2018MNRAS.475..676S} {475, 676}

\bibitem[\protect\citeauthoryear{{Stiskalek}, {Bartlett}, {Desmond}  \&
  {Anbajagane}}{{Stiskalek} et~al.}{2022}]{Stiskalek2022ML}
{Stiskalek} R.,  {Bartlett} D.~J.,  {Desmond} H.,   {Anbajagane} D.,  2022,
  \mn@doi [\mnras] {10.1093/mnras/stac1609}, \href
  {https://ui.adsabs.harvard.edu/abs/2022MNRAS.514.4026S} {514, 4026}

\bibitem[\protect\citeauthoryear{{Strickland} \& {Heckman}}{{Strickland} \&
  {Heckman}}{2009}]{Strickland2009SNEqns}
{Strickland} D.~K.,  {Heckman} T.~M.,  2009, \mn@doi [\apj]
  {10.1088/0004-637X/697/2/2030}, \href
  {https://ui.adsabs.harvard.edu/abs/2009ApJ...697.2030S} {697, 2030}

\bibitem[\protect\citeauthoryear{{Tissera}, {White}, {Pedrosa}  \&
  {Scannapieco}}{{Tissera} et~al.}{2010}]{Tissera2010BaryonImprints}
{Tissera} P.~B.,  {White} S. D.~M.,  {Pedrosa} S.,   {Scannapieco} C.,  2010,
  \mn@doi [\mnras] {10.1111/j.1365-2966.2010.16777.x}, \href
  {https://ui.adsabs.harvard.edu/abs/2010MNRAS.406..922T} {406, 922}

\bibitem[\protect\citeauthoryear{{Tr{\"o}ster} et~al.,}{{Tr{\"o}ster}
  et~al.}{2022}]{Troster2022KidstSZ}
{Tr{\"o}ster} T.,  et~al., 2022, \mn@doi [\aap] {10.1051/0004-6361/202142197},
  \href {https://ui.adsabs.harvard.edu/abs/2022A&A...660A..27T} {660, A27}

\bibitem[\protect\citeauthoryear{{Van der Walt}, {Colbert}  \&
  {Varoquaux}}{{Van der Walt} et~al.}{2011}]{vanderWalt2011Numpy}
{Van der Walt} S.,  {Colbert} S.~C.,   {Varoquaux} G.,  2011, \mn@doi
  [Computing in Science and Engineering] {10.1109/MCSE.2011.37}, \href
  {https://ui.adsabs.harvard.edu/abs/2011CSE....13b..22V} {13, 22}

\bibitem[\protect\citeauthoryear{{Vikram}, {Lidz}  \& {Jain}}{{Vikram}
  et~al.}{2017}]{Vikram2017GalaxyGroupstSZ}
{Vikram} V.,  {Lidz} A.,   {Jain} B.,  2017, \mn@doi [\mnras]
  {10.1093/mnras/stw3311}, \href
  {https://ui.adsabs.harvard.edu/abs/2017MNRAS.467.2315V} {467, 2315}

\bibitem[\protect\citeauthoryear{{Villaescusa-Navarro}
  et~al.,}{{Villaescusa-Navarro} et~al.}{2021}]{Villaescusa-Navarro2021CAMELS}
{Villaescusa-Navarro} F.,  et~al., 2021, \mn@doi [\apj]
  {10.3847/1538-4357/abf7ba}, \href
  {https://ui.adsabs.harvard.edu/abs/2021ApJ...915...71V} {915, 71}

\bibitem[\protect\citeauthoryear{{Virtanen} et~al.,}{{Virtanen}
  et~al.}{2020}]{Virtanen2020Scipy}
{Virtanen} P.,  et~al., 2020, \mn@doi [Nature Methods]
  {https://doi.org/10.1038/s41592-019-0686-2}, \href {https://rdcu.be/b08Wh}
  {17, 261}

\bibitem[\protect\citeauthoryear{{Weinberg}, {Mortonson}, {Eisenstein},
  {Hirata}, {Riess}  \& {Rozo}}{{Weinberg} et~al.}{2013}]{Weinberg:2013review}
{Weinberg} D.~H.,  {Mortonson} M.~J.,  {Eisenstein} D.~J.,  {Hirata} C.,
  {Riess} A.~G.,   {Rozo} E.,  2013, \mn@doi [\physrep]
  {10.1016/j.physrep.2013.05.001}, \href
  {https://ui.adsabs.harvard.edu/abs/2013PhR...530...87W} {530, 87}

\bibitem[\protect\citeauthoryear{{Weinberger} et~al.,}{{Weinberger}
  et~al.}{2017}]{Weinberger2017Methods}
{Weinberger} R.,  et~al., 2017, \mn@doi [\mnras] {10.1093/mnras/stw2944}, \href
  {https://ui.adsabs.harvard.edu/abs/2017MNRAS.465.3291W} {465, 3291}

\bibitem[\protect\citeauthoryear{{Weinberger} et~al.,}{{Weinberger}
  et~al.}{2018}]{Weinberger2018SMBHsIllustrisTNG}
{Weinberger} R.,  et~al., 2018, \mn@doi [\mnras] {10.1093/mnras/sty1733}, \href
  {https://ui.adsabs.harvard.edu/abs/2018MNRAS.479.4056W} {479, 4056}

\bibitem[\protect\citeauthoryear{{Wetzel}, {Hopkins}, {Kim},
  {Faucher-Gigu{\`e}re}, {Kere{\v{s}}}  \& {Quataert}}{{Wetzel}
  et~al.}{2016}]{Wetzel2016DwarfGals}
{Wetzel} A.~R.,  {Hopkins} P.~F.,  {Kim} J.-h.,  {Faucher-Gigu{\`e}re} C.-A.,
  {Kere{\v{s}}} D.,   {Quataert} E.,  2016, \mn@doi [\apjl]
  {10.3847/2041-8205/827/2/L23}, \href
  {https://ui.adsabs.harvard.edu/abs/2016ApJ...827L..23W} {827, L23}

\bibitem[\protect\citeauthoryear{{Wheeler} et~al.,}{{Wheeler}
  et~al.}{2019}]{Wheeler2019FIREdwarfs}
{Wheeler} C.,  et~al., 2019, \mn@doi [\mnras] {10.1093/mnras/stz2887}, \href
  {https://ui.adsabs.harvard.edu/abs/2019MNRAS.490.4447W} {490, 4447}

\bibitem[\protect\citeauthoryear{{York} et~al.,}{{York}
  et~al.}{2000}]{York2000SDSS}
{York} D.~G.,  et~al., 2000, \mn@doi [\aj] {10.1086/301513}, \href
  {https://ui.adsabs.harvard.edu/abs/2000AJ....120.1579Y} {120, 1579}

\bibitem[\protect\citeauthoryear{{Zacharegkas} et~al.,}{{Zacharegkas}
  et~al.}{2021}]{Zacharegkas2021GGLensingDES}
{Zacharegkas} G.,  et~al., 2021, arXiv e-prints, \href
  {https://ui.adsabs.harvard.edu/abs/2021arXiv210608438Z} {p. arXiv:2106.08438}

\bibitem[\protect\citeauthoryear{{van den Bosch}, {Ogiya}, {Hahn}  \&
  {Burkert}}{{van den Bosch} et~al.}{2018}]{VanDenBosch2018Subhalo}
{van den Bosch} F.~C.,  {Ogiya} G.,  {Hahn} O.,   {Burkert} A.,  2018, \mn@doi
  [\mnras] {10.1093/mnras/stx2956}, \href
  {https://ui.adsabs.harvard.edu/abs/2018MNRAS.474.3043V} {474, 3043}

\makeatother
\end{thebibliography}



\appendix

\section{Outlier rejection} \label{sec:Outlier}

The concentration, $\cvir$, for halos $\Mvir > 10^{11}\msol/h$ at a given redshift has a notable fraction of outliers with $\cvir \lesssim 1$. For these objects the scale radius is equivalent to $\Rvir$, and this is not a physical population of halos. We do not explore the exact reasons underlying the existence of this population, as it is beyond the scope of our work. However, we must remove such outliers before computing our scaling relations as they will bias the mean concentration to lower values. Thus, we require an adequate rejection algorithm.

\begin{figure}
    \centering
    \includegraphics[width = 1\columnwidth]{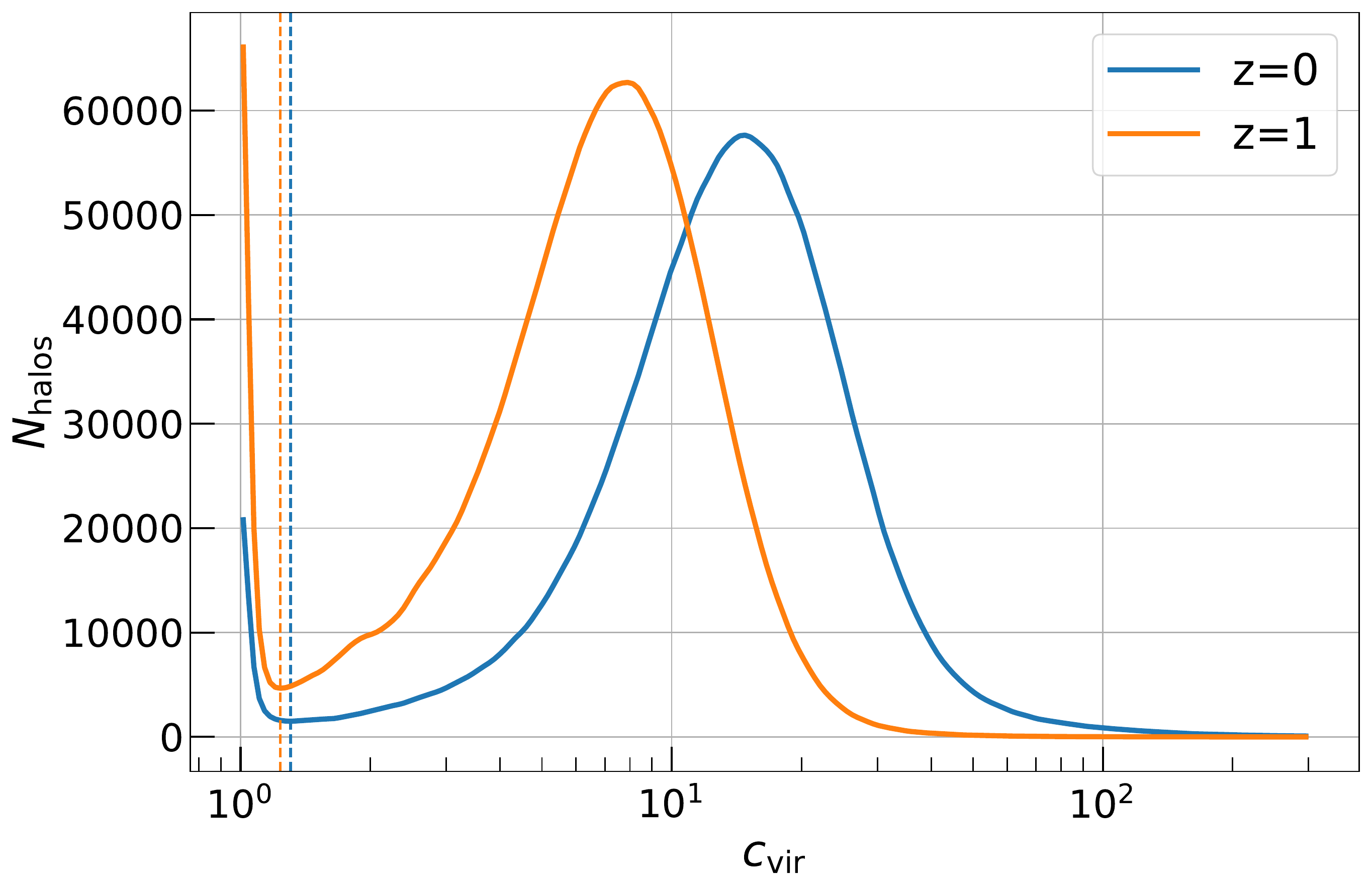}
    \caption{The distribution of $\cvir$ at $z = 0$ and $z = 1$, for halos with $\Mvir > 10^{11} \msol/h$. We see the distributions are generally well-behaved but have a sharp increase at low values of $\cvir \lesssim 1$. The vertical dashed lines indicate the inflection point of the distribution, which is where the histogram rises to the left to form the low $\cvir$ peak. All halos with $\cvir$ below this threshold are rejected.}
    \label{fig:OutlierMask}
\end{figure}

To reject outliers within a given halo sample, we first create a histogram of the halo concentration in 200 bins. A gaussian filter with a smoothing scale equal to the bin width then smooths the histogram to remove any noise fluctuations. We interpolate the function via a cubic spline. The interpolated distributions for $z \in \{0, 1\}$ for TNG are shown in Figure \ref{fig:OutlierMask}, with a clear peak at low $\cvir$. We use the inflection point at low $\cvir$ as our cutoff; all halos with $\cvir$ below this point are thrown out of the analysis. This inflection point is obtained by getting the first root of the derivative of the interpolated histogram. For $z = 0$ ($z = 1$), we find that $\approx 2\%$ ($\approx 6\%$) of halos are rejected using this procedure.

\section{Redshift dependence in SIMBA} \label{appx:SIMBA}

\begin{figure*}
    \centering
    \includegraphics[width =2 \columnwidth]{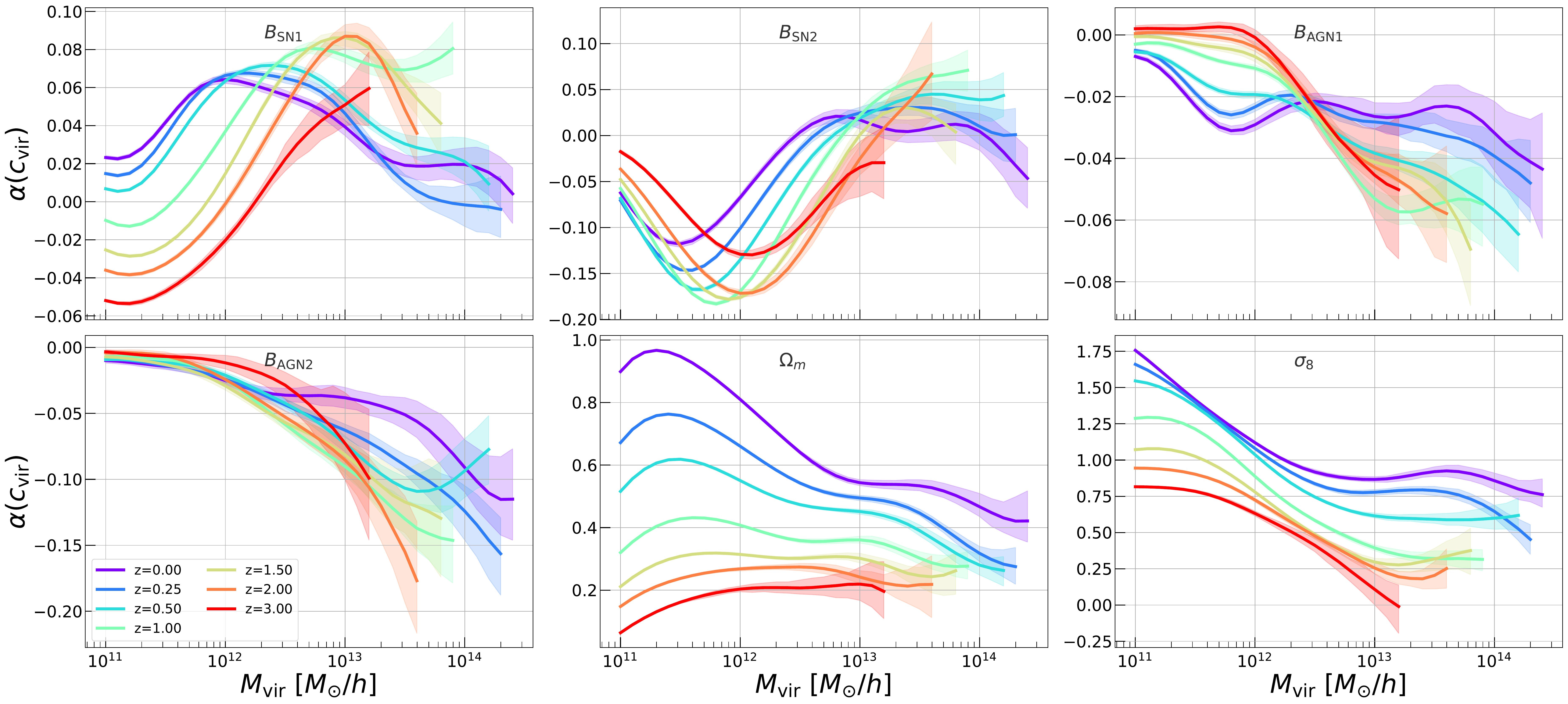}
    \caption{The \textsc{Kllr} slopes between halo concentration $c_{\rm vir}$ and the six astrophysical and cosmological parameters. Same format as Figure \ref{fig:slopes_z_dep}. The bands show the 68\% confidence interval from the model uncertainty. Similar to the case in \textsc{IllustrisTNG}, there is a clear and strong mass-dependence in all parameters, and a significant redshift evolution in most.}
    \label{fig:SIMBASlopes}
\end{figure*}

\begin{figure*}
    \centering
    \includegraphics[width =2 \columnwidth]{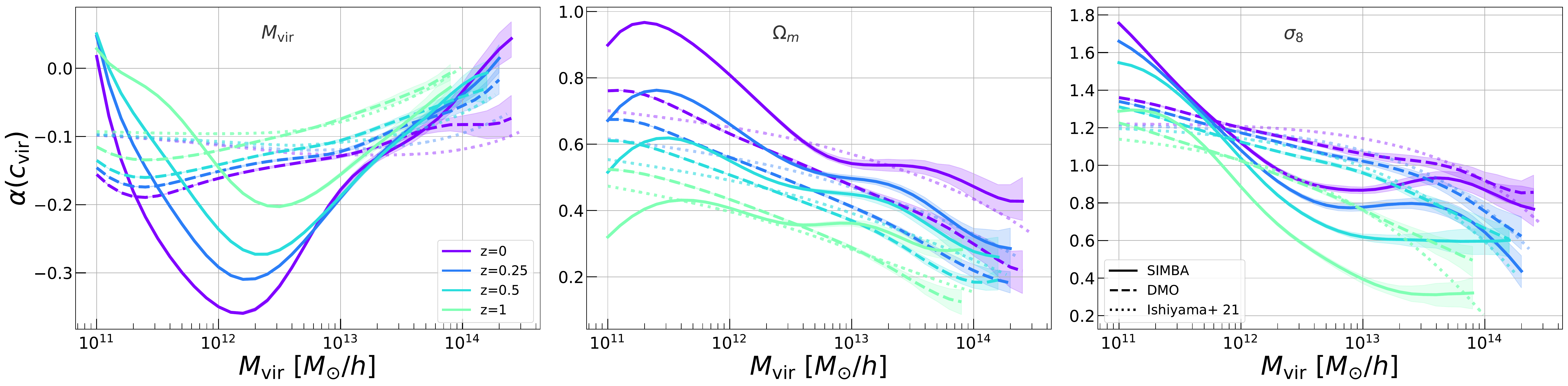}
    \hspace{15pt}
    \caption{The \textsc{Kllr} slopes of halo concentration $\cvir$ with halo mass $\Mvir$, $\Omega_{\rm m}$ and $\sigma_8$ at $0 < z < 1$. The format is the same as Figure \ref{fig:TNG_DMO_Colossus}. We compare the SIMBA slopes (solid) with the slopes from the DMO simulations (dashed) and predictions from the model of \citet{Ishiyama2020UchuuConcentration} using the \textsc{Colossus} package \citep{Diemer2018COLOSSUS}.}
    \label{fig:SIMBAandDMO}
\end{figure*}

The redshift dependence discussed in this work has so far been about the \textsc{IllustrisTNG} model (Figure \ref{fig:slopes_z_dep}); a choice made for brevity sake. In this appendix, we extract the redshift dependence in \textsc{Simba} and discuss it. Once again, we find that the slopes depend on mass in a non-monotonic way across all redshifts.

\textbf{In $\boldsymbol{B_{\rm SN1}}$}, the slopes are generally low in amplitude across all masses. There is a local maxima around $10^{12} \msol/h$ and the mass-scale of this maxima increases with redshift. The fact that the slopes are overall positive is unsurprising --- $B_{\rm SN1}$ controls the wind mass-loading factor which, as we have discussed previously, impacts the gas cooling efficiency through metal enrichment. Increases in this efficiency increase $\cvir$, leading to the generally positive slope, especially at lower redshifts where cooling has had more time to impact the matter distribution.

\textbf{In $\boldsymbol{B_{\rm SN2}}$}, we find behavior similar to $A_{\rm SN2}$; there is a minima in the slopes at $10^{11.5} \msol/h$ that increases with redshift. This is also unsurprising as both $A_{\rm SN1}$ and $B_{\rm SN2}$ control the SN wind speeds of their respective galaxy formation models, so even though the wind speeds are implemented differently in either model (i.e. different equations) both parameters change the same underlying quantity.

\textbf{In $\boldsymbol{B_{\rm AGN1}}$ and $\boldsymbol{B_{\rm AGN2}}$}, the slopes deviate more from 0 compared to the TNG counterparts. However, the amplitude of the slope is still quite weak. $B_{\rm AGN2}$ controls the jet speed, which is qualitatively similar to the ``burstiness'' controlled by $A_{\rm AGN2}$ and so we see similar behavior of slopes being zero at low mass but becoming non-zero at high mass. In fact the AGN slopes in \textsc{Simba} have a higher amplitude than their TNG counterpart.

\textbf{Finally, $\boldsymbol{\Omega_{\rm m}}$ and $\boldsymbol{\sigma_8}$} have similar trends in \textsc{Simba} as they did in \textsc{IllustrisTNG}. The dependence on $\sigma_8$ is quite similar, showing lower variation with redshift at low masses and stronger variations at high mass. The slopes in \textsc{Simba} seem to keep rising towards the lower masses, especially at low redshift, whereas in TNG they asymptoted to a flat, mass-independent trend. The redshift dependence of the $\Omega_{\rm m}$ slopes differs from that of TNG. At high masses there appears to be a constant multiplicative factor accounting for the redshift dependence. At low masses, the mass-dependence of the slope steepes more strongly at low redshift, and flattens at high redshift. Even though these are cosmological parameters that are detached from galaxy formation, the actual impact of cosmology on structure formation is some complicated convolution of cosmological evolution with galaxy formation. Given \textsc{Simba} and \textsc{IllustrisTNG} have different galaxy formation models, the dependence of $\cvir$ on cosmology will vary as well.

Lastly, for completeness, we also show in Figure \ref{fig:SIMBAandDMO} the redshift-dependent comparison of slopes from SIMBA and DMO runs, similar to Figure \ref{fig:TNG_DMO_Colossus}. The behavior in these plots follows from the discussions above. In general, there continue to be notable differences between the DMO and hydrodynamic runs, as is expected.

\bsp	
\label{lastpage}
\end{document}